\documentclass[useAMS,usenatbib]{mn2e}
\usepackage{amsmath}
\usepackage{amssymb}
\usepackage{multirow}
\usepackage{graphicx}
\usepackage{epstopdf}

\def\simlt{\lower.5ex\hbox{$\; \buildrel < \over \sim \;$}}
\def\simgt{\lower.5ex\hbox{$\; \buildrel > \over \sim \;$}}

\newcommand{\bd}{\begin{displaymath}}
\newcommand{\ed}{\end{displaymath}}
\newcommand{\be}{\begin{equation}}
\newcommand{\ee}{\end{equation}}
\newcommand{\beqa}{\begin{eqnarray}}
\newcommand{\eeqa}{\end{eqnarray}}

\title[Jetted TDEs as a Flag of IMBHs at High Redshifts] {Jetted Tidal Disruptions of Stars as a Flag of Intermediate Mass Black Holes at High Redshifts}
\author[Fialkov  \& Loeb] {Anastasia
  Fialkov$^{1}$\thanks{E-mail: anastasia.fialkov@cfa.harvard.edu},
  Abraham Loeb$^{1}$\\
  $^{1}$ Institute for Theory and Computation, Harvard University, 
  60 Garden Street, Cambridge, MA, 02138, USA\\
  }

\begin{document}
\pagerange{\pageref{firstpage}--\pageref{lastpage}} \pubyear{2015}
\maketitle

\label{firstpage}

\begin{abstract}

Tidal disruption events (TDEs) of stars by single or binary supermassive black holes (SMBHs) brighten galactic nuclei and reveal a population of otherwise dormant black holes. Adopting event rates from the literature, we aim to establish general trends in the redshift evolution of the TDE number counts and their observable signals. We pay particular attention to  (i) jetted TDEs whose luminosity is boosted by relativistic beaming, and  (ii) TDEs around binary black holes. We show that  the brightest (jetted) TDEs are expected to be produced by massive black hole binaries if the occupancy of  intermediate mass black holes (IMBHs) in low mass galaxies is high and if the TDE luminosity is proportional to the black hole mass.  The same binary population will also provide gravitational wave sources for eLISA. In addition, we find that the  shape of the X-ray luminosity function of TDEs strongly depends on the occupancy of IMBHs and could be used to constrain scenarios of SMBH formation. Finally, we make predictions for the expected number of TDEs observed by future  X-ray telescopes finding that a 50 times more sensitive instrument than the  Burst Alert Telescope (BAT) on board the  {\it Swift} satellite is expected to trigger  $\sim 10$  times more events than BAT, while  6-20 TDEs are expected in each deep field observed by a telescope 50 times more sensitive than the {\it Chandra X-ray Observatory} if the occupation fraction of IMBHs is high. Because of their long decay times, high-redshift TDEs can be mistaken for fixed point sources  in deep field surveys  and targeted observations of the same deep field with year-long intervals could reveal TDEs.

\end{abstract}

\begin{keywords}
cosmology: theory--galaxies: jets--X-rays: bursts
\end{keywords}

\section{Introduction}
\label{Sec:Intro}

Supermassive black holes (SMBHs) with masses between $10^6$ and  $10^{10}~M_\odot$ \citep{Ghisellini:2010, Thomas:2016} are observed to reside at the centers  of dark matter  halos with masses $\gtrsim 10^{12}~M_\odot$. Smaller dark matter halos, such as hosts of present day dwarf galaxies or  galaxies at high redshifts, are expected to be populated with  intermediate mass black holes  (IMBHs) with masses in the range $\sim 10^2- 10^6~M_\odot$ \citep{Greene:2012}. The origin of SMBHs and IMBHs is still not well understood. In hierarchical structure formation these black holes are expected to grow from initial seeds as a result of galaxy mergers in which black holes coalesce   \citep[see][for review]{Graham:2015}. Gas-rich mergers fuel AGN emitting energy in the optical, ultraviolet and X-ray bands. Roughly 10\% of low-redshift AGN ($z\lesssim 5$) are radio-loud \citep{Jiang:2007},  producing a pair of collimated relativistic jets which could be observed to greater distances because of the relativistic beaming effect. The fraction of radio-loud quasars at higher redshifts ($z\sim6$) was shown to be  $8^{+5.0}_{-3.2}$\% \citep{Banados:2015},  suggesting no evolution of the radio-loud fraction with $z$.  On the other hand, if the merger is dry and the merging galaxies do not have enough gas to feed the black hole, a dormant massive black hole (MBH) results without  observable electromagnetic signature. 

Even though the samples of both SMBHs and IMBHs build up, the percentage of galaxies hosting central  black holes (the so-called occupation fraction)  is still unclear, especially in low mass galaxies \citep{Greene:2007}, and until recently IMBHs were considered hypothetical. However, recent observations have shown that some of dwarf galaxies in the local Universe could indeed be populated by IMBHs \citep{Farrell:2009, Reines:2013, Moran:2014, BALDASSARE:2015, BALDASSARE:2016}. The growing observational evidence includes 151 dwarf galaxies with candidate black holes in the mass range $10^5-10^6~M_\odot$ as identified from optical emission line ratios and/or broad H$\alpha$  emission \citep{Reines:2013};  28 active galactic nuclei (AGN) with black hole masses in  $10^3- 10^4~M_\odot$ range in nearby low-mass, low-luminosity dwarf galaxies found with the Sloan Digital Sky Survey  \citep{Moran:2014}. In addition,  \citet{Lemons:2015} showed that large fraction of hard X-ray sources in dwarf galaxies are ultra-luminous, suggesting that they actually are IMBHs;  \citet{Yuan:2014} describe four dwarf Seyferts with masses $< 10^6~M_\odot$; \citet{BALDASSARE:2015} reported observations of a $5\times 10^4~M_\odot$ black hole in the dwarf galaxy RGG 118; while  \citet{BALDASSARE:2016} list eleven additional black holes with masses between $7\times 10^4  - 1\times 10^6~M_\odot$. Finally, it was shown that more than  $20$\% of early-type galaxies with a stellar mass less that $ 10^{10}~M_\odot$ are expected to have massive black holes  in their cores \citep{Miller:2015}. All this observational evidence suggests that IMBHs in dwarf galaxies are not as exotic as previously thought. However, dynamical mass measurements suggest a decline in the occupation fraction of black holes in host galaxies with velocity dispersion below 40 km s$^{-1}$ \citet{Stone:2016KO}.

One way to improve our constraints on the black hole occupation fraction is by probing the population of quiescent black holes when they are temporary illuminated by a tidal disruption event (TDE) in which a star passing too close to the black hole is shredded by a gravitational tide which exceeds the self-gravity of the star. Theoretical work on TDEs spans several decades, including works by  \citet{Hill:1975, Frank:1976,  Lacy:1982, Carter:1983, Rees:1988, Evans:1989, Phinney:1989, Magorrian:1999, Wang:2004, Perets:2006, Guillochon:2013, Stone:2013, Stone:2016, Roth:2016} and others. When a TDE occurs, part of the stellar mass is ejected away, forming an elongated stream and heating the ambient medium  \citep{Guillochon:2016}, while the bound debris are accreted by the black hole emitting bright observable flare at a wide range of wavelengths from radio to  $\gamma$-rays \citep{Rees:1990}. However, for very massive black holes \citep[$\sim 1\times 10^8 M_\odot$ for a solar mass star][]{Kesden:2012} the tidal disruption distance is smaller than the Schwarzschild radius  and stars are swallowed whole without exhibiting TDE flares. The emission peaks in the UV or soft X-rays with typical peak luminosity  in the  soft X-rays band being $L_{X}\sim 10^{42}-10^{44}$ erg s$^{-1}$. The flare decays on the timescale of months or years as a power law with a typical index $-5/3$, which is often considered to be the telltale signature of a tidal disruption of a star by a massive black hole. 

Following the first detection by ROSAT \citep{Bade:1996, Komossa:1999},   about 50 TDEs have been observed \citep{Komossa:2015} in hard X-ray \citep{Bloom:2011, Burrows:2011, Cenko:2012, Pasham:2015}, soft X-ray \citep{Bade:1996, Komossa:1999, Donley:2002, Esquej:2008, Maksym:2010, Saxton:2012, Saxton:2016}, UV  \citep{Stern:2004, Gezari:2006, Gezari:2008, Gezari:2009} and optical \citep{vanVelzen:2011, Gezari:2012,  Arcavi:2014, Chornock:2014, Holoien:2014, Vinko:2015} wavelengths. Some of the observed TDEs exhibit unusual properties. In particular, one of the detected TDEs shows an excess of variability in its light curve \citep{Saxton:2012} which can be explained if the black hole is actually a binary with a mass of $10^6~M_\odot$, mass ratio of 0.1 and semimajor axis of 0.6 milliparsecs \citep{Liu:2014}. This candidate appears to have one of the most compact orbits among the known SMBH binaries and has overcome the ``final parsec problem'' \citep{Colpi:2014}. Upon coalescence, it will be  a strong source of gravitational wave emission in the sensitivity range of eLISA. Three other TDEs appear to be very bright in X-rays  with peak soft X-ray isotropic luminosity being highly super-Eddington \citep{Burrows:2011, Cenko:2012, Brown:2015}, while followup observations  showed that these events were also associated with bright, compact, variable radio synchrotron emission \citep{Zauderer:2011, Cenko:2012}.  The observed high X-ray luminosity can be explained if the tidal disruption of stars in these cases  powered a highly beamed relativistic jet pointed at the observer  \citep{Tchekhovskoy:2014}. Based on these three observations, \citet{Kawamuro:2016} concluded that  $0.0007\% - 34\%$ of all TDE source relativistic jets, while \citet{Bower:2013} and \citet{vanVelzen:2013} estimate that $\lesssim 10\%$ of TDEs produce jetted emission at the observed level. Formation of jets in TDE is a topic of active research, e.g., works by   \citet{Metzger:2011, Mimica:2015, Generozov:2016}.

Based on the observations the TDE rate was derived to be  $10^{-4}-10^{-5}$ per year per galaxy \citep{Donley:2002, Khabibullin:2014, Esquej:2008, Luo:2008, Maksym:2010, Gezari:2008, Wang:2012, vanVelzen:2014} and is consistent  with order-of magnitude theoretical predictions when IMBHs are ignored. The TDE rates were shown to be sensitive to the density profile and relaxation processes taking place in galactic nuclei \citep{Magorrian:1999, Wang:2004, Stone:2016}. The simplest and most commonly used estimate of the TDE rates is based on the steady-state solution of the Fokker-Planck equation describing the diffusion of stars in angular momentum and energy space driven by two-body relaxation. This process re-populates stellar orbits along which stars are disrupted  by MBH, the so-called ``loss cone''.  With the two-body relaxation being the main process to refill the loss cone, other processes that may contribute to the stellar budget were also discussed in literature \citep{Rauch:1996, Hopman:2006, Merritt:2015, Bar-Or:2016, Perets:2006, Magorrian:1999,  Merritt:2004, Vasiliev:2013, Vasiliev:2014, Chen:2009, Li:2015, Wegg:2011, Liu:2013, Ivanov:2005, Chen:2011, Merritt:2005, Lezhnin:2015, Lezhnin:2016}.

It is still unclear why IMBHs do not contribute to the observed TDEs, and most of the related theoretical studies show that black holes with masses smaller than $10^6~M_\odot$ can disrupt stars at rates higher than those of higher masses \citep{Wang:2004,Stone:2016}. Therefore, if  small halos are occupied by IMBHs,  most disruptions are expected to occur in these systems  making TDEs particularly good probes of  the poorly-understood, low-mass end of MBH mass function \citep{Stone:2016}. Moreover, once detected in large quantities, TDEs will offer insight into physics of quiescent  black holes, probe extreme accretion physics near the last stable orbit, provide the means to measure the spin of black holes and probe general relativity in the strong-field limit \citep{Hayasaki:2016, Guillochon:2015c}. In addition, jetted TDEs will allow us to explore processes through which relativistic  jets are born.
 
In this paper we extrapolate the population of TDEs to high redshifts (out to $z=20$), and predict their detectability with the  next-generation X-ray telescopes. We propose a new way to test the occupation fraction of IMBHs through their unique contribution to the X-ray luminosity function. The paper is organized as follows: We summarize our approach  in Section \ref{Sec:Methods}, deriving TDE rates as a function and outlining the TDE luminosity prescriptons. We  showing the intrinsic X-ray luminosity function in Section \ref{Sec:LF}. Next, we make predictions for realistic next generation X-ray surveys in Section \ref{Sec:Obs} focusing on upgrades of {\it Swift} and {\it Chandra}. We summarize our conclusions in Section \ref{Sec:sum}.

\section{Model Components}
\label{Sec:Methods}

Even though TDEs have been extensively studied, the predicted rates are not in good agreement with observations. Therefore, we adopt simple assumptions for the event rates from literature with the aim to establish general trends in the redshift evolution of the TDE number counts and their observable signals. After defining the population of galaxies and black holes in Section \ref{Sec:BH}, we start by considering rates in a given galaxy of halo mass  $M_h$ (Section \ref{Sec:TDE})  and generalize for a cosmological population of galaxies in Section \ref{Sec:Cosm}. We discuss the luminosity of TDE flares in Section \ref{Sec:Lum}.

\subsection{Galaxies and Black Holes}
\label{Sec:BH}
One of the key model ingredients  that determines the  TDE rates is the distribution of stars in galactic nuclei  \citep{Magorrian:1999, Wang:2004}. Depending on the merger history of the galaxy and the efficiency of feedback on star formation, the stellar density profile can develop either  a core or a cusp. For simplicity we adopt a singular isothermal sphere (SIS)  density profile $\rho(r) = \sigma^2/2\pi G R^2$ with $\sigma$ being the constant velocity dispersion and  $R$ the halo virial radius.  For a galaxy of halo mass $M_h$, the relation between  the halo mass and the velocity dispersion is simply $M_h =2\sigma^2R /G$; while the velocity dispersion can be directly related to the black hole mass using the  $M_{BH}-\sigma$ relation \citep{Kormendy:2013, McConnell:2013, Saglia:2016, BALDASSARE:2015, Thomas:2016}
 \begin{equation}
 M_{BH} =  0.309\times 10^9\times \left(\sigma /200~\textrm{km s$^{-1}$}\right)^{4.38}~M_\odot,
 \label{Eq:Msigma}
 \end{equation}
which holds for a wide range of black hole masses  from  $5\times 10^4$ M$_\odot$ \citep{BALDASSARE:2015} to $1.7\times 10^{10}$ M$_\odot$ \citep{Thomas:2016} in galaxies with a bulge \citep{Guillochon:2015b}. Assuming the isothermal stellar distribution, \citet{Wang:2004} derived TDE rates for galaxies with a single central black hole, while \citet{Chen:2009} report the rates in a case of a black hole binary. As we discuss in  Section \ref{Sec:TDE}, for MBH with masses in the range $M_{BH}\sim 10^5-10^{8}~M_{\odot}$ the TDE rates per halo computed using the isothermal stellar distribution are similar (within tens of percent) to more realistic estimates based on a large galaxy sample \citep{Stone:2016}, which justifies our assumption.  The error in the rate estimation due to the idealized stellar density profile is small compared to other uncertainties, e.g., introduced by the poorly constrained occupation fraction of IMBHs in low mass galaxies, which amounts to one-two orders of magnitude uncertainty in the derived volumetric TDE rates. 

In order to address the uncertainty in the occupation fraction, $f_{occ}$, of  MBH  we consider two extreme cases:  (i) complete black hole occupation of all halos that form stars, and (ii) assume that there are no black holes with masses below $10^6~M_{\odot}$, which is equivalent to the vanishing occupation fraction  in halos below $10^{10}-10^{11}~M_\odot$ (depending on redshift). We refer to the former case as MBHs (or  $f_{occ}=1$) and latter case as SMBHs (or $f_{occ}=0$). The two cases can be considered as an upper (former case) and lower (latter case) limits for the occupation fraction yielding, respectively, upper and lower limits for the expected TDE rates. 

Even though black hole seeds could exist in ever smaller halos in the hierarchical picture of structure formation, one also needs stars to fuel a TDE flare. The lowest mass of a halo in which stars can form at high redshifts before the end of hydrogen reionization at $z\sim 6$ is determined by the cooling condition, which involves either molecular or atomic hydrogen \citep{Tegmark:1997, LoebFurlanetto:2013, Barkana:2016}. The lowest temperature coolant, molecular hydrogen, forms stars in dark matter halos as tiny as $\sim 10^5 ~M_\odot$. However, hydrogen molecules are easily destroyed by radiative feedback  \citep{Machacek:2001} in which case star formation proceeds via atomic cooling in halos of $10^7-10^8~M_{\odot}$. Here we neglect the  molecular cooling channel and assume that before reionization galaxies can form in halos down to a velocity dispersion of $\sim 12$ km s$^{-1}$, which host black holes of mass $10^{3.1}~M_{\odot}$. After reionization is complete, the smallest star forming halos are sterilized by photoheating feedback which evaporates gas out of all halos with velocity dispersion less than $\sim 25$    km s$^{-1}$. As a result, small galaxies stop producing many stars and the loss cone of stars around the central black hole is most likely not refilled efficiently enough to support the equilibrium TDE rates. Therefore, we assume in the post-reionization era that all black holes below $\sim 10^{4.5}~M_{\odot}$ remain without fuel and do not source TDEs. In a realistic reionization scenario, the minimal halo mass which efficiently forms stars would gradually rise with redshift \citep{Bacchic:2013, Cohen:2016}. However, because the reionization history is poorly constrained at present,  we adopt the simplest scenario of instantaneous reionization at $z_{re}=8$, consistent with latest constraints by the {\it Planck} satellite \citep{Planck:2016}. The minimal black hole mass  in our MBHs scenario is thus
\[M_{BH,min}  = \left\{
  \begin{array}{lr}
    10^{3.1}~M_{\odot}, & z \ge 8\\
    10^{4.5}~M_{\odot}, & z < 8
  \end{array}
\right.
\]

In our second, conservative, scenario we assume that $M_{BH, min} =10^6~M_\odot$. Several effects can contribute to low TDE rates from IMBHs, justifying our SMBHs scenario: black holes could be kicked out of halos as a result of merger; radiation from AGN could have negative feedback on star formation (AGN feedback), in this case the loss cone would not be replenished.  Another possible feedback mechanism is the stellar feedback from supernova explosions which can expel gas from the halo making star formation less efficient \citep{Wyithe:2013}. 

An additional model ingredient that determines the TDE rate is the merger history of a halo which we incorporate in Section \ref{Sec:Cosm}. As discussed in Section \ref{Sec:TDE}, the  TDE rate in a recently merged galaxy is boosted by several orders of magnitude for $\sim 10^5$ years compared to a galaxy with a quiet merger history, e.g., works by  \citet{Ivanov:2005, Chen:2009, Chen:2011}. The enhanced TDE rates are explained by the fact that the dynamics of the system are changed by the presence of a black hole binary produced  as a result of a merger. 

\subsection{TDE rates per halo}
\label{Sec:TDE}

In the case of a single black hole, the most secure way to feed stars into the loss cone around the black hole is via the standard two-body relaxation, which  sets a lower limit on the TDE rates between $10^{-4}$ and $10^{-6}$ yr$^{-1}$ \citep{Frank:1976, Lightman:1977, Cohn:1978, Magorrian:1999, Wang:2004, Stone:2016}. Other processes   that may contribute to the stellar budget include resonant relaxation \citep{Rauch:1996, Hopman:2006, Merritt:2015, Bar-Or:2016}, presence of massive perturbers  such as stellar clusters or gas clouds \citep{Perets:2006}, nonspherical geometry \citep{Magorrian:1999,  Merritt:2004, Vasiliev:2013, Vasiliev:2014}, black hole binaries \citep{Chen:2009, Li:2015, Wegg:2011, Liu:2013, Ivanov:2005, Chen:2011}, and anisotropy in the initial conditions \citep{Merritt:2005, Lezhnin:2015, Lezhnin:2016}. 

Assuming that the loss cone is refilled via two-body relaxation, the rate of tidal disruptions per halo per year  for an isothermal stellar distribution \citet{Wang:2004}  reads
 \begin{equation}
\dot N^{1h}_{TDE} \sim 2.47\times 10^{-4} \left(\frac{\sigma}{100~\textrm{km s$^{-1}$}}\right)^{7/2}\left(\frac{M_{BH}}{10^7~M_{\odot}}\right)^{-1}~\textrm{yr$^{-1}$}
\label{Eq:TDE1}
\end{equation}
where we only considered the disruption of solar mass stars\footnote{The dependence of Eq. (\ref{Eq:TDE1}) on stellar mass can be re-introduced by adding a factor  $\left(m_*/M_{\odot}\right)^{-1/3}\left(r_*/R_{\odot}\right)^{1/4}$ with $r_* = R_{\odot}(m_*/M_{\odot})^{0.8}$ for stars along the lower main sequence.}.  Despite the fact that the stellar density used to derive Eq. (\ref{Eq:TDE1}) is idealized, the rates are similar to those derived by \citet{Stone:2016} for stellar profiles from a real galaxy sample. \citet{Stone:2016} estimated TDE rates due to two-body relaxation from $\sim 200$ galaxies  with $M_{BH}\sim 10^5-10^{10}~M_{\odot}$ at $z\sim 0$ and got  $\dot N_{TDE}\sim 1.2\times 10^{-5}\left(M_{BH}/10^8 M_\odot \right)^{-0.247}$ for galaxies with a core and $\dot N_{TDE}\sim 6.5 \times 10^{-5}\left(M_{BH}/10^8 M_{\odot}\right)^{-0.223}$ for galaxies with a cusp.  We checked that for our choice of $M_{BH}-\sigma$ relation Eq. (\ref{Eq:TDE1}) gives similar rates   in normalization (up to several tens of percents) and comparable slope of $\sim -0.2$ when compared to the cusp fit of \citet{Stone:2016}.

Modeling TDE rates in merging system is more challenging, and rates are less understood than disruptions by a single MBH. Even though all theoretical studies point in the direction of TDE rates boosted for $\sim 10^5$ years by $\sim 2$ orders of magnitude compared to the  disruption by single MBH, there is no consensus on details and it is unclear at present what is the leading process that  replenishes the loss cone.  When two galaxies merge, the black holes in the galactic cores first inspiral toward each other due to  the  dynamical friction. Next, when the mass in gas and stars enclosed within the black hole orbit is smaller than the total mass of the two black holes, the black holes become gravitationally bound and evolve as a binary. For black hole masses of $\sim 10^6~M_{\odot}$ this occurs when the typical separation between the two black holes is $\sim$parsec. Gradually, the binary hardens. If the binary reaches separations $\lesssim 0.001$ pc, gravitational waves are emitted as the two MBHs coalesce. Each one of the stages in the evolution of the binary has its own rate of TDEs. Using N-body simulations to model  dry major mergers, \citet{Li:2015} conclude that in the first stage,  the tidal disruption rate for two well separated MBHs in merging system has similar levels to the sum of the rates of two individual MBHs in two isolated galaxy. In their fiducial model \citet{Li:2015} find that  after two MBHs get close enough to form a bound binary,   the disruption rate is  enhanced by   a factor of 80 within a short time lasting for 13 Myr. This boosted disruption stage finishes after the SMBH binary evolves to a compact binary system, corresponding to a drop back of the disruption rate to a level few times higher than for an individual MBH. Other studies also point in the direction of enhanced rates from binaries. In particular, \citet{Ivanov:2005} considered  secular evolution of stellar orbits in the gravitational potential of an unequal mass binary and derived rates of $10^{-2}-1$ TDEs per year per galaxy for a $10^6-10^7$ M$_\odot$ primary black hole and a binary mass ratio $q> 0.01$. The duration of this boosted disruption stage was determined by the dynamical friction time scale  
\begin{equation}
T_{dyn}\sim \frac{2\times 10^2(1+q)}{q} \left(\frac{10^7~ \textrm{M}_\odot}{\textrm{M}_{\textrm{BH}}}\right)^{1/2} ~\textrm{yr}.
\end{equation}
\citet{Chen:2009} used scattering experiments to show that gravitational slingshot interactions between hardened binaries and a bound spherical isothermal stellar cusp will be accompanied by a burst of TDEs. It appears that a significant fraction of stars initially bound to the primary black hole will be scattered into the loss cone by resonant interaction with the secondary black hole. \citet{Chen:2009} provide a fitting formula for the maximal TDE rates per halo with a binary MBH system  
\begin{equation}
\dot N^{2h}_{TDE} \sim (1+q)^{1/2} \left(\frac{\sigma}{100~\textrm{km s$^{-1}$}}\right)^4\left(\frac{M_{BH}}{10^7 ~M_\odot}\right)^{-1/3} ~\textrm{yr}^{-1}.
\label{Eq:TDE2}
\end{equation} 
and show that the enhancement lasts for $\sim 10^4$ years. \citet{Chen:2011} included the Kozai-Lidov effect, chaotic three-body orbits, the evolution of the binary and the non-Keplerian stellar potentials and  found  that for masses of $ 10^7$ M$_\odot$ and $10^5$ M$_\odot$, TDE rates  0.2 events per year last for $\sim 3\times 10^5$ years which is three orders of magnitude larger than for a single black hole and broadly agrees with the conclusions of \citet{Chen:2009}. \citet{Wegg:2011} included the same processes as \citet{Chen:2011} and arrived at similar rates. They found that the majority of TDEs are due to chaotic orbits in agreement with \citet{Chen:2009}, showing that  the  Kozai-Lidov effect plays a secondary role. Their rates are somewhat smaller than in \citet{Chen:2009} largely because the authors consider less cuspy stellar profiles.  \citet{LiNaoz:2015} considered the evolution of stellar disruption around a binary with MBHs masses of 10$^7$ M$_\odot$ and 10$^8$ M$_\odot$ due to the eccentric Kozai-Lidov mechanism yielding  rate of 10$^{-2}$ TDE per year for $5\times 10^5$ years. Finally, \citet{Liu:2013} concluded that the TDE rates of stars by SMBHs in the early phase of galaxy merger when galactic dynamical friction dominates could also be enhanced by several orders of magnitude (up to $10^{-2}$ events per year per galaxy) as a result of the perturbation by companion galactic core and the triaxial  gravitational potential of the galactic nucleus. 
 
To accommodate tidal disruptions induced by binary MBHs we adopt the fitting function given by Eq. (\ref{Eq:TDE2}) assuming  that this enhanced  rate last for a dynamical time $T_{dyn}$, while for the rest of the time the TDE rate is simply $\dot N^{1h}_{TDE}$. As we can derive from Eqs. (\ref{Eq:TDE1})  and (\ref{Eq:TDE2}) the scaling of TDE rates with $M_{BH}$  is different for single and binary MBHs. Applying the $M_{BH}-\sigma$ relation to Eqs. (\ref{Eq:TDE1})  and (\ref{Eq:TDE2}) we find that the TDE rate scales as $\dot N^{2h}_{TDE}\propto M_{BH}^{0.6}$ for binaries and  $\dot N^{1h}_{TDE} \propto M_{BH}^{-0.2}$ for single black holes. This property has immediate implications to the total observable TDE rates that will be discussed in Section \ref{Sec:Obs}.

\subsection{Number of TDEs across cosmic time}
\label{Sec:Cosm}

The observed TDE number counts per unit time that originate from redshift $z$ depends on several factors with the dominant factor being the amount of halos of each mass and their merger history. To determine the halo abundance we make use of the Sheth-Tormen mass function \citep{Sheth:1999} in calculating the comoving number density of halos in each mass bin $dN_h/dM_h$ in units of $M_{\odot}^{-1}$ Mpc$^{-3}$. Next, adopting the merger rates of \citet{Fakhouri:2010} we calculate the dimensionless average merger rate $dN_m/d\zeta /dz$ (in units of mergers per halo per unit redshift per unit  halo mass ratio, $\zeta$), given by a fitting formula 
\begin{equation}
\frac{dN_m}{d\zeta dz}(\textrm{M},\zeta,z) =A\left(\frac{\textrm{M}}{10^{12}\textrm{M}_{\odot}}\right)^{\alpha} \zeta^{\beta}\exp\left[\left(\frac{\zeta}{\bar \zeta}\right)^{\gamma}\right](1+z)^{\eta}, 
\end{equation}
where $(\alpha,\beta,\gamma,\eta) = (0.133,-1.995,0.263,0.0993$ and $(A,\bar \zeta)=(0.0104,9.72\times10^{-3})$. 

\begin{figure}
\centering
\includegraphics[width=3.4in]{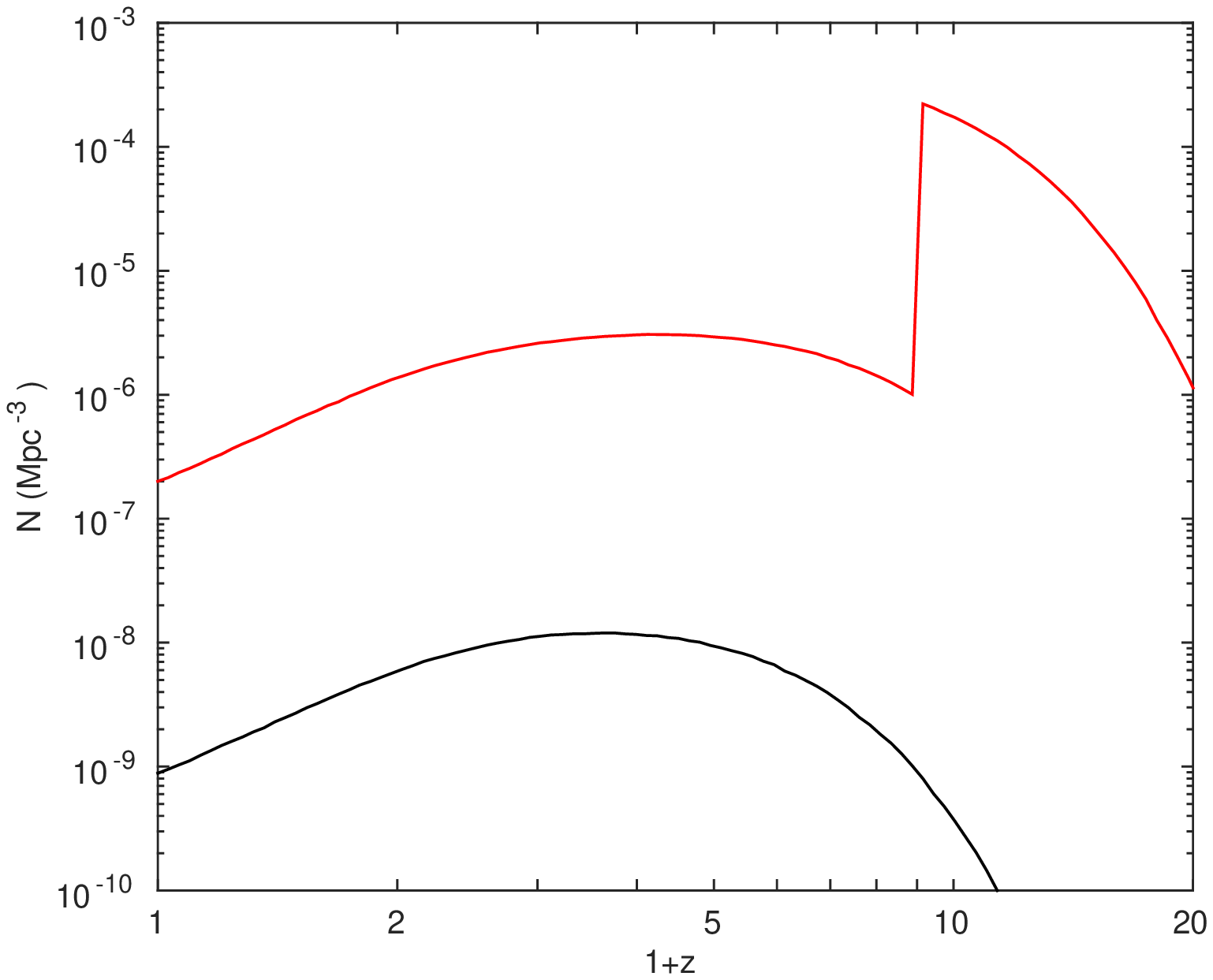}
\includegraphics[width=3.4in]{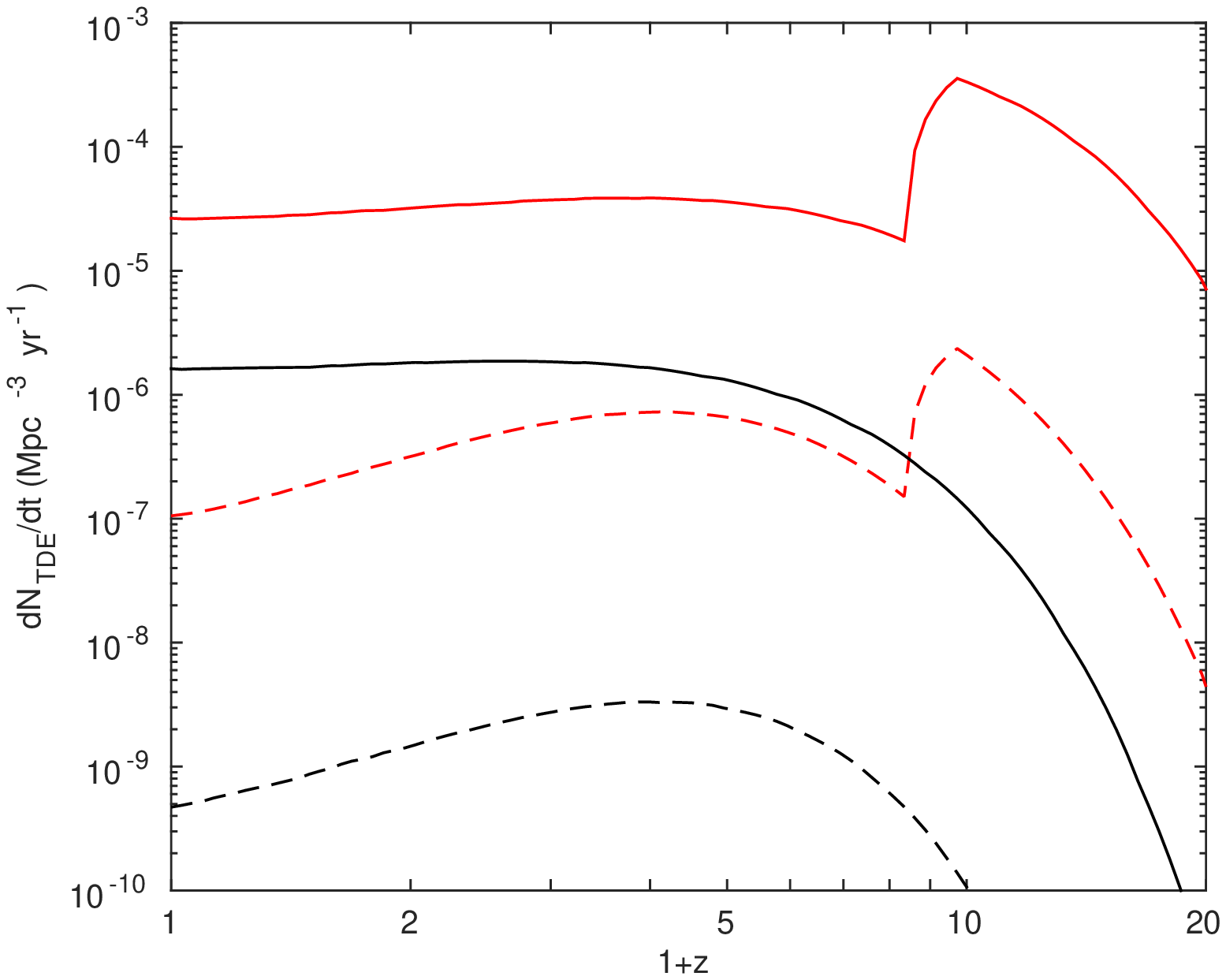}%\includegraphics[width=3.4in]{Nm}
\caption{Top: Halo mass averaged number density of systems with mergers that occurred at $t<T_{dyn}$. We show the case of SMBHs only ($f_{occ} = 0$, black) and all MBHs ($f_{occ} = 1$, red).  Bottom: Intrinsic TDE rates  in the observer's frame per comoving volume as a function of redshift are shown for SMBHs only (black) and all MBHs (red). In each case we show the contribution of single black holes with rates from \citet{Wang:2004} (solid, $\dot N^{1h}_{TDE}$) and contribution of binaries (dashed, $\dot N^{2h}_{TDE}$) assuming solar mass stars.  In all cases with mergers we assume that the enhancement due to binaries lasts for $T_{dyn}$.}
\label{Fig:1}
\end{figure}

For each halo we  assign a TDE rate of $\dot N^{2h}_{TDE}$ according to the probability, $P_{2}$, of it to encounter  a recent merger, and  $\dot N^{1h}_{TDE}$ with a probability $P_1 = 1-P_{2}$. The probability, $P_2$, is determined using the following criterion: if the time between mergers is larger than the dynamical time,  the TDE rates are those of single MBH, while if the time between mergers is smaller than $T_{dyn}$, there is an enhancement due to  binaries. Given the merger rates, we estimate the probability of a halo of mass $M_h$ at redshift $z$ to be a result of a recent merger as follows
\begin{displaymath}
P_{2}(M_h,z) = 1-\exp\left[- T_{dyn}\lambda\right],
\end{displaymath}
where 
\begin{equation}
\lambda = \int d \zeta \frac{dN_m}{dz d\zeta}\frac{dz}{dt}.
 \label{Eq:lambda}
\end{equation}
For very light halos, mergers are frequent and halos typically undergo several mergers within $T_{dyn}$, in which case the probability for a merger is near unity. We assume that probability for merger with black hole mass ratio $q$ is flat for $q = 10^{-3}-10^{-1}$, and $q$ is related to the halo mass ratio, $\zeta$, through Eq. (\ref{Eq:Msigma}). We ignore mergers with $0.1<q<1$ as they are expected to be rare. The top panel of Figure \ref{Fig:1} shows the halo mass weighted number density of mergers $\int dM_h  P_{2} dN_h/dM_h$ that yield enhanced TDE rates per unit volume in cases of  $f_{occ} = 0$ (SMBHs only) and  $f_{occ} = 1$ (all MBHs). In each case, the integral is over halos whose progenitors have both large enough black holes and gas to form stars. Such mergers are rare when the minimal mass is high (SMBH case), especially at high redshifts. In the case of $f_{occ}=1$ we can clearly see the turn on of photoheating feedback  at $z=8.8$ which shuts down star formation in galaxies below $M_h\sim 10^9~M_{\odot}$ at lower redshifts leading to suppressed TDE rate.  

Having the proper probabilities we can now calculate the expected TDE rates for an observer as a function of redshift. At every given redshift $P_{1}$ halos in each mass bin host a single MBH yielding TDE rate  
\begin{equation}
\dot N^{1}_{TDE} = \int dM_h \frac{dN_h}{dM_h}P_{1}\frac{ \dot N^{1h}_{TDE}}{(1+z)}.
\end{equation}
The factor $(1+z)$ compensates for the time dilation in the apparent rate. The contribution from binaries is given by 
\begin{equation}
\dot N^{2}_{TDE} = \int dM_h \frac{dN_h}{dM}  \int d q \frac{dP_{2}}{dq} \frac{\dot N^{2h}_{TDE}}{(1+z)}.
\end{equation}
The total number density of TDE per year per unit comoving volume, $\dot N_{TDE} = \dot N^{1}_{TDE}+\dot N^{2}_{TDE}$, is shown  on the bottom panel of Figure \ref{Fig:1}. 

\begin{figure}
\centering
\includegraphics[width=3.4in]{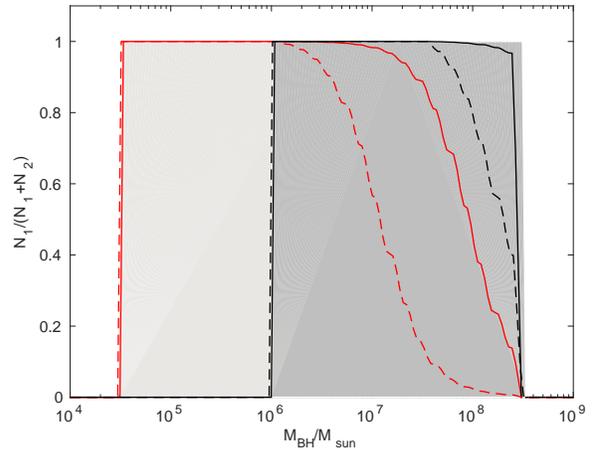}%\includegraphics[width=3.4in]{Rates0b}
\caption{Fraction of intrinsic disruption events sourced by single black holes versus black hole mass for SMBHs only ($f_{occ} = 0$,  black curves) and all MBHs ($f_{occ} = 1$, red curves). We show the fraction at $z = 0$ (solid) and $z=5$ (dashed). The dark grey region marks the occupation of SMBH, whereas the pale grey refers to the occupation of IMBHs.} 
%Expected number of jetted TDEs in a shell of depth $dz = 0.2 z$ assuming that 10\% of TDEs source jets with $\Gamma = 10$, $\theta = 1/\Gamma$. Each event injects two jets and the fraction of sky covered by one event is thus $2\pi \theta^2/4\pi $. Same color code is used. }
\label{Fig:2}
\end{figure}

As was pointed out above, TDE rates induced by binary black holes are higher in the high black hole mass end, while the single black hole systems are more efficient in the low black hole mass  end. To demonstrate this feature, we show on Figure \ref{Fig:2}  the fraction of intrinsic events (with no (1+z) factor)  sourced by single black holes out of total number of TDEs at $z=0$ and $z=5$ as a function of the black hole mass (total mass in the case of binaries) in solar mass units for $f_{occ} =0$ and 1. As expected from the TDE scaling with the black hole mass ($\propto M_{BH}^{0.6}$ for binaries and $\propto M_{BH}^{-0.2}$ for single black holes), binary systems dominate at large black hole masses and at high redshifts (because of the increased merger rates). The mass dependence determines contribution of each component to the overall luminosity function of the TDEs which we consider in the next section. 

Finally, in our cosmological model we  assume that 10\% of all TDE source jets with the Lorenz factor  of order $\Gamma = 10$ based on X-ray observations of jetted TDE \citep{Burrows:2011}. In the spirit of our simple approach we ignore parameters, such as stellar magnetic fields \citep{Guillochon:2016}, which could affect fraction of jetted TDEs and assume a constant jet fraction over the halo mass and redshift range. Because the luminosity is channeled into a collimated beam of an opening angle   $\theta \sim 1/\Gamma$, only a small fraction of the jetted TDEs will be actually observable. For an observer,  the fraction of sky covered by the jets pointing toward the observer  is, thus,  $ f_{jet} = 10\%\times 2\times \pi \theta^2/4\pi =5\times 10^{-2}$\% where we accounted for two jets emitted by every system. Overall, the observed TDE rates  will include  90\% of non-jetted TDE  and  $ f_{jet}$ of TDE with jets where we account only for the events that point toward the observer. 

\subsection{Luminosity of TDE flares}
\label{Sec:Lum}
The TDE rates shown in Figure \ref{Fig:1} are not the ones we would actually observe. Observable rates depend on the luminosity (observed flux) of each event as well as on the sensitivity of a telescope (discussed in the next Section). In this section we outline our assumptions for the TDE luminosity and use them in Section \ref{Sec:Obs} to calculate the observable signals. Because we are mainly interested in the high-redshift events which could be observed in X-rays when relativistic jets are produced,  we will focus on the TDE signature in X-rays ignoring their UV and optical counterparts. We are also largely ignoring radio signals because it is unclear whether or not the radio emission from the TDE sourced jets arises from the same regions as their X-ray emission.  The three X-ray observations of TDEs with jets include: (1) SwiftJ164449.3+573451, hereafter Sw1644+57 \citep{Bloom:2011, Burrows:2011,Levan:2011},  of peak X-ray isotropic luminosity  $L_X\sim 4\times 10^{48}$ erg s$^{-1}$ which originated from a galaxy at  $z=0.353$ and  was  discovered in March 2011 by the {\it Swift} Burst Alert Telescope \citep[BAT, 15-150 keV, ][]{Barthelmy:2005}; (2) SwiftJ2058.4+0516, hereafter SwJ2058+05 \citep{Cenko:2012}, of peak X-ray isotropic luminosity equivalent to  $L_X\sim 4\times 10^{48}$ erg s$^{-1}$, which   was discovered at $z=1.185$ in May 2011 by the BAT as part of the hard X-ray transient search; and (3) SwiftJ1112.2-8238, Sw1112-82 hereafter \citep{Brown:2015}, which was detected by BAT in June 2011 as an unknown, long-lived gamma-ray transient source  in a host identified at $z=0.89$ and with  $L_X\sim 6\times 10^{48}$ erg s$^{-1}$. Estimates of the SMBH mass in each one of these events yield  $M_{BH}\sim 10^6- 10^7~M_\odot$. Because the Eddington luminosity for a black hole of mass $M_{BH}$ is only $L_{Edd} = 1.3\times 10^{38} \left( M_{BH}/M_{\odot}\right)$ erg s$^{-1}$, these events either are intrinsically highly super-Eddington or the emitted energy is channeled in tightly collimated jets and the luminosity is boosted by a factor of $\sim 10^3-10^4$. 

Theory predicts that a flare is produced when debris return to the vicinity of a black hole $t_{fall} \approx 0.1 \left(M_{BH}/10^6~M_\odot\right)^{1/2}$  years after the disruption  and forms an accretion disc. If a solar mass star is completely disrupted, its debris fallback rate is  \citep{Rees:1988, Phinney:1989, Stone:2013}
\begin{equation}
\dot M_{fall}\sim \frac{1}{3t_{fall}}\left(\frac{t}{t_{fall}}\right)^{-5/3}~M_\odot~\textrm{yr$^{-1}$}
\label{Eq:Mfall}
\end{equation}
with the peak mass accretion rate value of
\begin{equation}
\frac{\dot M_{peak}}{\dot M_{Edd}} = 133 \left(\frac{M}{10^6~M_\odot}\right)^{-3/2},
\label{Eq:Mac}
\end{equation}
where the Eddington accretion rate is $\dot M_{Edd} = L_{Edd}/\eta c^2$ and $\eta \sim 0.1$ is a typical radiative efficiency. It is likely that the mass fallback rate can be directly related to the observed X-ray luminosity of the source and, thus, can be used to determine the total emitted energy. In particular, if the accretion rate is fully  translated to the bolometric luminosity,  the peak luminosity is $L_{peak} = \eta c^2 \dot M_{peak}$.  However, it is still not clear what is the efficiency of this process especially for intermediate black holes with mass less than 50 million solar masses for which $L_{peak}$ is highly super-Eddington for efficient  circularization of the debris  \citep{Guillochon:2015, Dai:2015, Shiokawa:2015}. 

Super-Eddington accretion fueled by a tidal disruption of a star was both detected in Nature \citep{Kara:2016} and studied in numerical simulations \citep{McKinney:2014, Sadowski:2015a, Sadowski:2015b, Jiang:2014, Inayoshi:2016, Sakurai:2016}. Based on observations of reverberation in the redshifted iron K$\alpha$ line \citep{Kara:2016}, super-Eddington accretion  was detected in one of the jetted TDE events, Sw1644+57. From the reverberation timescale, the authors estimate the mass of the black hole to be a few million solar masses, suggesting an accretion rate of at least 100 times the Eddington limit. Simulations  suggest that relativistic beaming can explain observed super-Eddington luminosities indicating that,  once the accretion is super-Eddington, relativistic jets can be produced \citep{McKinney:2014}. Although the overall radiative efficiency and luminosity are still debated, in simulations a strong outflow is generated and radiation can leak through a narrow funnel along the polar direction. Close to the black hole, a jet carves out the inner accretion flow, exposing the X-ray emitting region of the disk. \citet{Sadowski:2015a} found that if a source with moderate accretion rate is observed down the funnel, the apparent luminosity of such a source will be orders of magnitude higher than the non-jetted luminosity. \citet{Sadowski:2015b} show that for an observer viewing down the axis, the isotropic equivalent luminosity  is as high as $10^{48}$ erg s$^{-1}$ for a $10^7~M_\odot$ black hole accreting at $10^3$ the Eddington rate, which agrees with the observations of jetted TDEs.  Independent of the accretion rate in simulations, super-Eddington disks around black holes exhibit a surprisingly large efficiency of $\eta \sim 4\%$ for non-rotating black holes; while spinning  black holes yield  the maximal efficiency of jets of 130\% \citep{Piran:2015}. 

In other simulated systems such as stellar black holes, observed super-Eddington luminosities are also inferred:   \citet{Jiang:2014}  studied super-Eddington accretion flows onto black holes using a global three dimensional radiation magneto-hydrodynamical simulation and found  mass accretion rate of $\dot M \sim 220 L_{Edd}/c^2$  with outflows along the rotation axis, and radiative luminosity of $10 L_{Edd}$; $\dot M \sim 400 L_{Edd}/c^2$ was measured for a $10~M_\odot$ black hole with peak luminosity of $50 L_{Edd}$ \citep{McKinney:2015};   \citet{Inayoshi:2016} argued  that $\dot M \sim 5000 L_{Edd}/c^2$ is limited to the  Eddington luminosity in a metal poor environment, but \citet{Sakurai:2016} find $1 < L/L_{Edd} < 100$.

Because the TDE flares in jetted events are not fully understood \citep[although see][]{Crumley:2016}, we use two simple approaches to derive first the bolometric, and then the X-ray, luminosity for each event. Our first approach (Model A) is to simply assign the Eddington luminosity to each event according to the black hole mass, $L_{TDE}^A = L_{Edd}$. The second approach (Model B) assumes that the TDE bolometric luminosity is proportional to the mass accretion rate. However, for IMBHs the peak accretion rate exceeds  both the observed rates and the simulated ones by few orders of magnitude. As studies have shown,  TDE luminosity is not likely to exceed few hundreds $L_{Edd}$ \citep{McKinney:2014, Sadowski:2015a, Sadowski:2015b, Jiang:2014, Inayoshi:2016, Sakurai:2016}. Therefore, we adopt an upper limit of $300 L_{Edd}$ and the luminosity of each event reads 
\begin{equation}
L_{TDE}^B = \textrm{min}\left[L_{peak},300 L_{Edd}\right].
\label{Eq:Lacc}
\end{equation}
The major distinction between Models A and B is that in Model A the brightest events are produced by the biggest black holes which also are the rarest ones, especially at high redshifts; while when the luminosity scales as the accretion rate with a ceiling (Model B), the most luminous events are produced by black holes of mass $M_{BH}\sim 2.5\times 10^{6} ~M_{\odot}$ which are more common.

Observations show that the X-ray luminosity of the three jetted TDEs has a spectral energy distribution (SED) well fitted by a power law $S_\nu \propto \nu^{-\alpha}$ with a spectral index $\alpha$ in the range of $0.3-1$ with $\alpha  = 0.33$ for   Sw1112-82,  $\alpha \sim 0.8$  for Sw1644+57 and $\alpha \sim 0.6$ for  SwJ2058+05. Therefore, in our modeling we adopt power-law SED with a unique spectral index of $\alpha = 0.5$ to describe all the jetted events. The SED  of a non-jetted TDE is expected to be a combination of a power-law and a black body, where the latter is negligible at high enough energies \citep[$\sim 1$ keV and above,][]{Kawamuro:2016}. We follow \citet{Kawamuro:2016} assuming that the spectral index, $\alpha$, is the same for non-jetted events as for the jetted ones ($\alpha =0.5$). The intrinsic spectral luminosity of an event is thus $L_\nu = L_0\nu^{-\alpha}$ where $L_0$ is the normalization constant. Assuming that the SED of these objects over a wide wavelength range is similar to that of AGN, we can calculate the X-ray luminosity of each event based on its bolometric luminosity. For the soft X-ray band ($2-10$ keV) we adopt a bolometric correction factor of  $k_{2-10}\sim 50$ for the Eddington   and  $k_{2-10} = 70$ for the super-Eddington accretion rates    \citep{Kawamuro:2016, Vasudevan:2007}. Given these numbers we normalize our power law spectra in the soft X-ray band so that  $L_0 = k_{2-10}(1-\alpha)L_{TDE}\left[10^{1-\alpha}-2^{1-\alpha}\right]^{-1}$. Using this prescription we can calculate the  observed spectral flux for  non-jetted TDEs  at redshift $z $ 
\begin{equation}
S_\nu = \frac{L_0 \nu^{-\alpha} (1+z)^{1-\alpha}}{4\pi D_L^2},
\end{equation}
 where $D_L$ is the luminosity distance to the source. In a jet, the observed flux at an observed frequency $\nu$   is  boosted by the factor of $\mathcal{D }^{3+\alpha}$ where  $\mathcal{D }= \left[\Gamma (1-B\mu_{obs})\right]^{-1}$ is the Doppler factor and $\mu_{obs} = \cos\theta$  is the angle of the jet with respect to the observer \citep{Burrows:2011}. 

Simulations show that TDEs occurring around MBH binaries have similar peak luminosity in X-rays as TDEs sourced by single black holes; however, the light curve has stronger variability in time due to the perturbations introduced by the secondary black hole \citep{Liu:2009, Liu:2014, Coughlin:2016, Ricarte:2016}. Therefore,  we adopt similar prescription as described above to  asigned X-ray luminosity to TDEs sourced by binaries.

\section{Intrinsic Luminosity Function}
\label{Sec:LF}

We can now make predictions for the intrinsic X-ray luminosity function of TDEs in the two cases of SMBHs and MBHs. It appears that, because  TDE rates from single and binary black holes scale differently with $M_{BH}$,  binaries dominate TDE production in the most massive halos and, as a result, contribute the brightest TDE flares in Model A. However, this contribution is significant only when small dark matter halos are occupied by IMBHs providing enough progenitors to form SMBH binaries. In the case of Model B, the most luminous events happen in systems with $M_{BH}\sim 2.5 \times 10^6~M_{\odot}$ which are dominated by single black holes. In case only SMBHs populate halos, TDEs from binaries occur only in systems with both black holes of $M_{BH}>10^6~M_\odot$  which are extremely rare and contribute at most  few percent of the brightest TDE flares.

\begin{figure}
\centering
\includegraphics[width=3.4in]{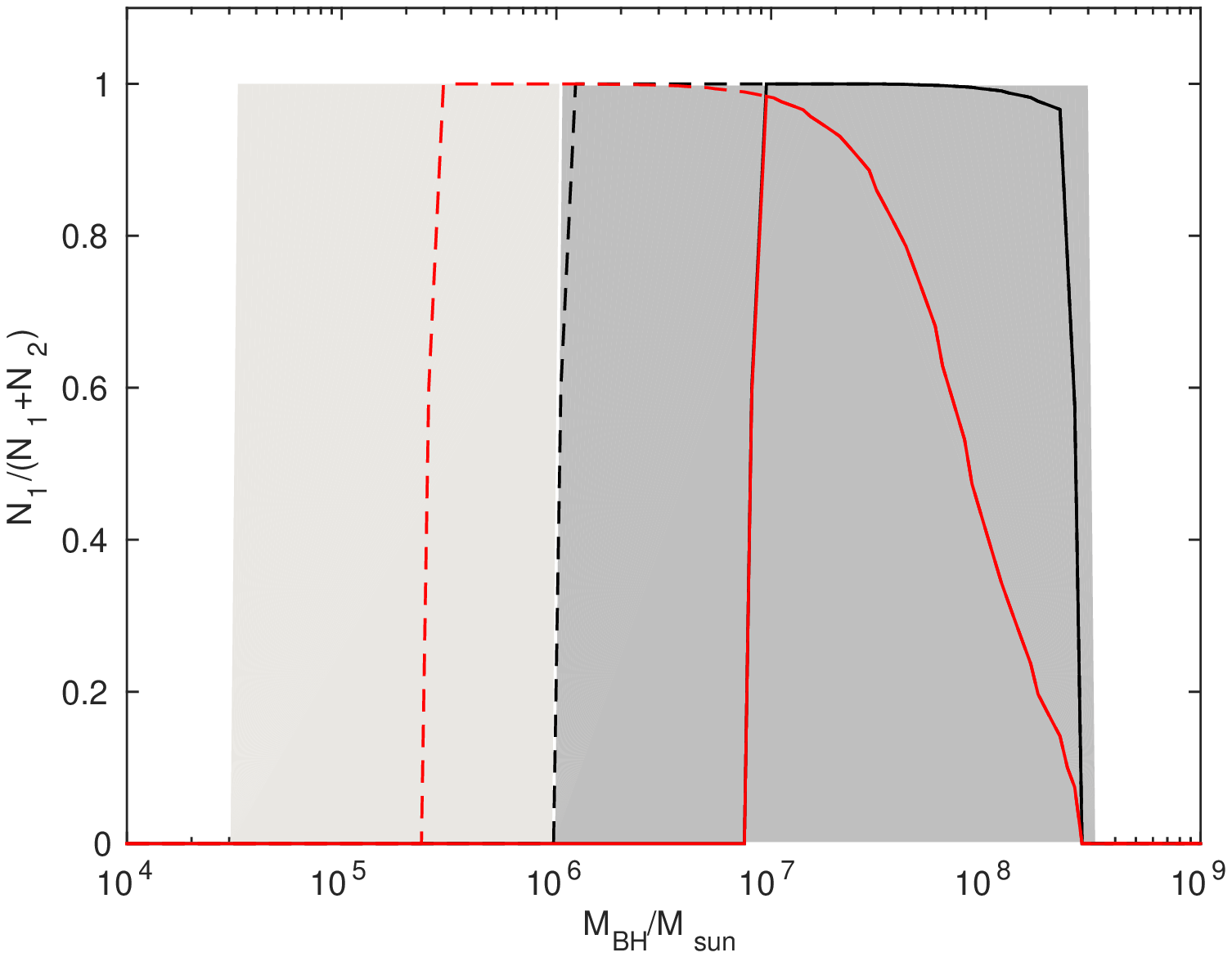}
\includegraphics[width=3.4in]{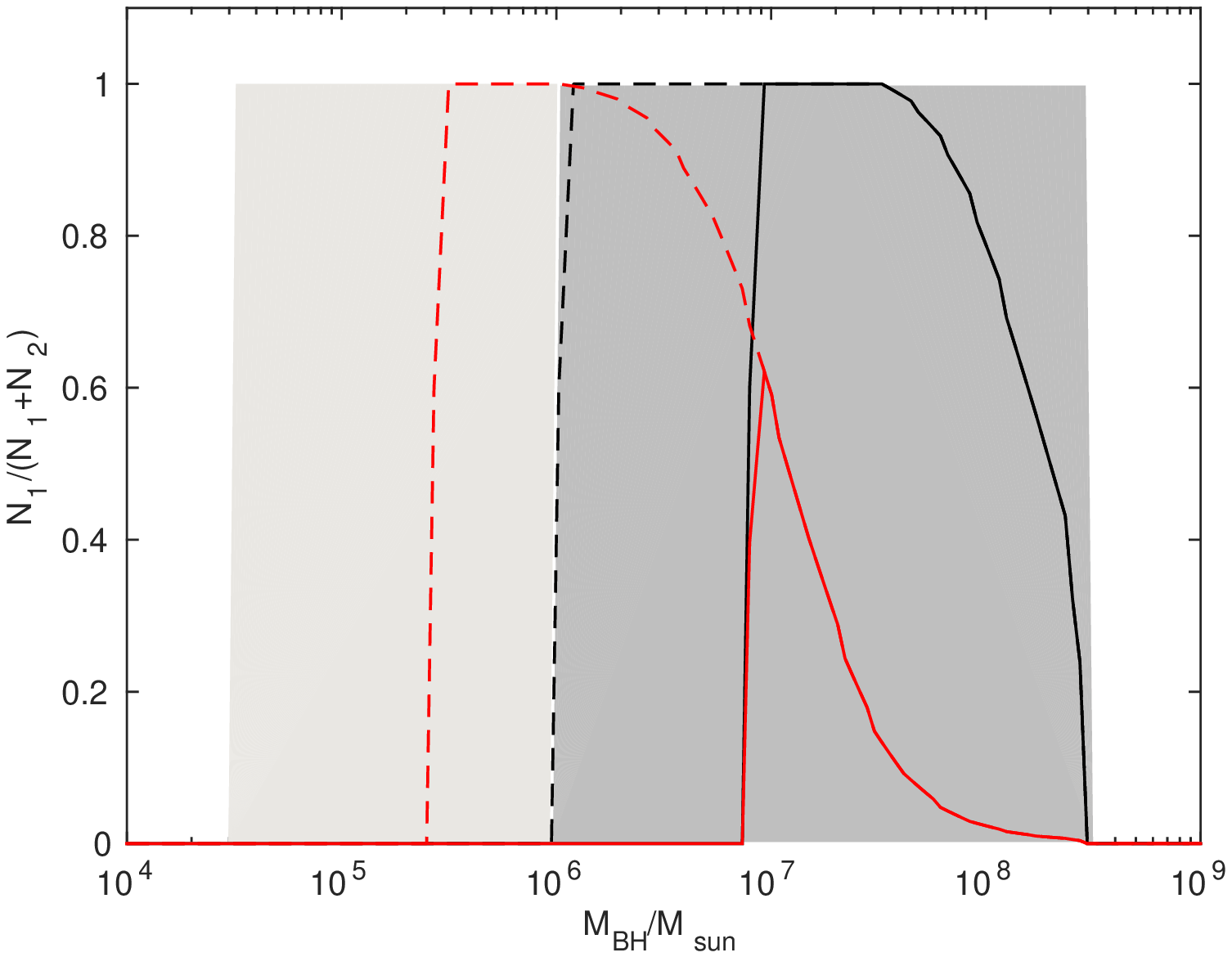}
\caption{Fraction of TDEs brighter than an intrinsic luminosity $L> 10^{45}$ erg s$^{-1}$ (close to the Eddington luminosity of $M_{BH}=10^7~M_\odot$) that are sourced by single MBHs, $N^{1}_{TDE}/(N^{1}_{TDE}+N^{2}_{TDE})$,  as a function of $M_{BH}$  at $z=0$ (top) add $z=5$ (bottom). We show the case of Model A (solid) and B (dashed) for SMBH  (black) and MBHs (red). The dark grey region marks the occupation of SMBH, with pale grey referring to IMBHs.}
\label{Fig:3}
\end{figure}

The fraction of events brighter than $ 10^{45}$ erg s$^{-1}$ (which is close  to the Eddington luminosity of a $M_{BH}=10^7~M_\odot$ black hole) is shown on Figure \ref{Fig:3}.  To reinforce this point, we list in Table \ref{Table:1} the percentage of events brighter than $10^{45}$ erg s$^{-1}$ produced by binary black holes at redshifts $z=$0, 0.5, 1, 2, 5, 7, 10, and 15. The importance  of binaries grows toward higher redshifts where mergers are more frequent. In the case of fully occupied halos, binaries start dominating  the bright events at $z=1$ in Model A and their contribution increases with redshift; while in Model B the maximal fraction of bright TDEs  sourced by binaries is only  $\sim 11\%$ in the post-reionization era ($z\lesssim 8$). In both Models A \& B with $f_{occ}=1$ we find a sudden increase in the binary contribution at $z>8$ (pre-reionization era) when the photoheating feedback is not active. In the case of $f_{occ} =0$, as expected,  the fraction of binaries is at most few percents and varies smoothly with redshift as this population is not affected by the photoheating feedback. With next generation X-ray telescopes  which could statistically analyze high redshift TDEs, the change in TDE number counts  with redshift could be a smoking gun of feedback processes or a marker  of the black hole occupation fraction.   

\begin{table}
		\begin{center}
			\begin{tabular}{llllllllll}
				\hline 
				 Model & Redshift & All &    Bright (A)&    Bright (B)   \\
		\hline   
				$f_{occ}=1$ & $z=0$ &   0.4\%  & 25.2\% & 2.1\%\\					
						& $z=0.5$ &   0.7\%  & 38.3\% & 3.6\%\\								
				            & $z=1$ &   1.0\%  & 50.8\% & 5.1\%\\				            
				            & $z=2$ &   1.6\%  & 66.7\% & 8.7\%\\				            				               
				            & $z=5$  &   1.5\%  & 83.8\% & 11.1\%\\				            				            
				            & $z=7$  &   0.9\% & 85.7\% & 8.3\%\\	
				            & $z=8$  &   0.9\% & 98.2\% & 68.6\%\\				            				            			            
				            & $z=10$  &   0.5\%  & 98.4\% & 69.8\% \\				            
				            & $z=15$  &   0.1\%  & 98.8\% & 74.4\% \\

\hline
				$f_{occ}=0$ & $z=0$   &  $<0.1$\% &   0.2\% & $<0.1$\%  \\										
				 & $z=0.5$   &  $<0.1$\%  &  0.3\% &  $<0.1$\%  \\						 	 
				 & $z=1$   &  $<0.1$\%  &  0.5\% & $<0.1$\%  \\				 
				 		   & $z=2$   &  0.2\%  & 0.9\% &    0.2\%  \\				 		   
						      &  $z=5$   &  0.2\%  & 2.2\% &   0.2\% \\						      				            
				            &  $z=7$   &  0.1\%  & 2.2\% &   0.1\%  \\	
				            &  $z=8$   &  0.1\%  & 2.2\% &   0.1\%  \\	
				              &  $z=10$   &  $<0.1$\% & 1.8\%  &  $<0.1$\%  \\
				            &  $z=15$   &  $<0.1$\%  & 0.5\%  &  $<0.1$\%  \\
				\hline
	\end{tabular}
				\caption{Percentage of  TDE  produced by binary black holes: all events (3rd column), only bright events with intrinsic luminosity $L> 10^{45}$ erg s$^{-1}$ for  Model A (4th column) and  Model B (5th column). }
				\label{Table:1}
\end{center}
\end{table}

\begin{figure}
\centering
\includegraphics[width=3.4in]{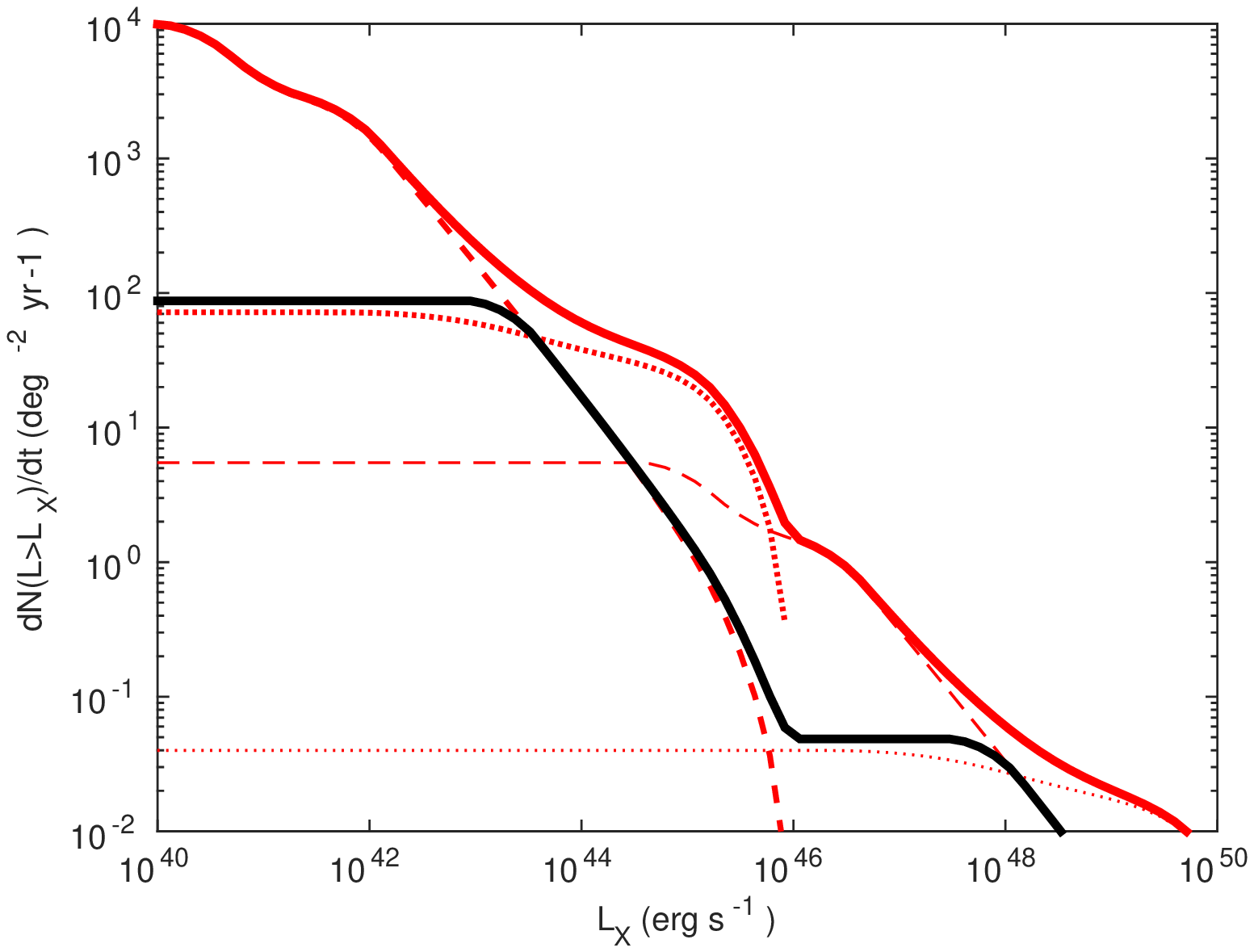}
\includegraphics[width=3.4in]{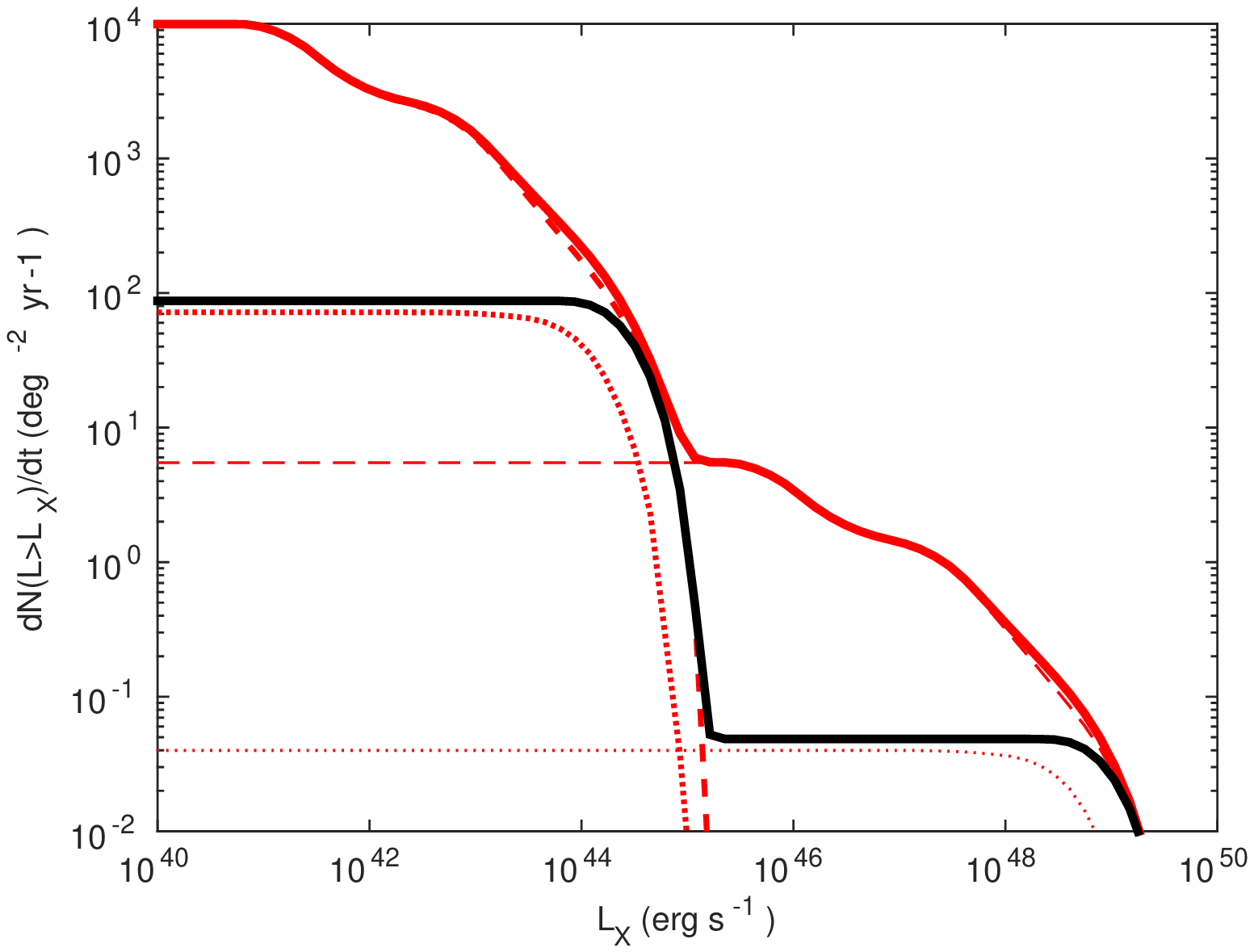}
\caption{Number of events per year detected by an ideal instrument of a field of view of 1 deg$^2$ versus X-ray luminosity in the observed 1-150 keV band.  We show the cumulative luminosity functions assuming sources with Eddington luminosity, i.e., Model A, (top) and luminosity that scales as the mass accretion rate, i.e., Model B, (bottom), with $f_{occ}=1$ (red, thick solid) and $f_{occ}=0$  (black, thick solid).  For the $f_{occ}=1$ case, we also show contributions due to various components:  non-jetted events sourced by   single black holes (thick dashed red) and binary black holes (thin dashed red), jetted events sourced by  single black holes (thick dotted red) and binary black holes (thin dotted red). The brightest events are dominated by jetted TDE sourced by binary black holes, but these are very rare. }
\label{Fig:4}
\end{figure}

We conclude this section by showing the expected cumulative X-ray luminosity function for Model A (top panel of Figure \ref{Fig:4}) and Model B (bottom panel of Figure \ref{Fig:4}) versus X-ray luminosity in the observed 1-150 keV band.  Our results are presented for both SMBHs and MBHs. In the case of MBHs there are four distinct terms (shown separately on the figure) that affect the luminosity function and contributes a ``knee'', i.e., the contributions from single and binary black holes of jetted and non-jetted events. When the TDE luminosity scales with the halo mass (Model A) each on eof these terms  dominates  specific luminosity regime, while the impact of binaries  is less evident in the case of Model B.  On the other hand, our SMBHs model has only two distinct features because in this case mergers have a negligible contribution and the knees in the luminosity function result from the jetted and non-jetted population of TDEs produced by single black holes. The shape of the luminosity function alone could be used to place limits on the occupancy of the IMBHs once a complete compilation  of TDEs is available.

\section{Observational signature}
\label{Sec:Obs}

Number of disruption events that are actually detected by a telescope depend on its flux limit and field of view. Here we will focus on telescopes such as  {\it Swift} and {\it Chandra} and explore signals which next-generation X-ray missions could probe. Bright X-ray transients such as  GRBs or jetted TDEs are  detected when they first trigger the BAT on {\it Swift}. The trigger occurs  if the signal's flux rises above $28.8\times 10^{-11}$ erg cm$^{-2} s^{-1}$ in the hard X-ray band (15-150 keV), i.e., reaches the 6-$\sigma$ statistical significance of BAT  \citep{Barthelmy:2005}.   Interestingly, all three jetted TDEs were  detected by {\it Swift} over a period of three consecutive months, which suggests the possibility that further examples may be uncovered by detailed searches of the BAT archives. The X-ray Telescope (XRT) is another instrument on board  {\it Swift} observing in the soft X-ray   band (0.2-10 keV) and reaching $2\times 10^{-14}$  erg cm$^{-2}$ s$^{-1}$ sensitivity in $10^4$ seconds with a $23.6\times 23.6$ arcmin$^2$ field of view. (Because  soft X-ray photons below $\sim 1$ keV can be absorbed by  dust, we will quote numbers in the observed $2-10$ keV   band when referring to soft X-rays.) As we show below, a telescope with such field of view and sensitivity as XRT is good for follow up observations of TDEs; while either a larger field of view or sensitivity are required to detect TDEs in large quantities.  In fact, a telescope such as {\it Chandra} with its high point source sensitivity of $\sim 4\times 10^{-15}$ erg cm$^{-2}$ s$^{-1}$ in $10^4$ s (or $\sim 4\times 10^{-17}$ erg cm$^{-2}$ s$^{-1}$ in $10^6$ s) over 0.4-6 keV band and field of view of $\sim 15\times 15$ arcmin$^2$, could have many TDEs per frame, as we argue below.

\begin{figure*}
\centering
\includegraphics[width=3.4in]{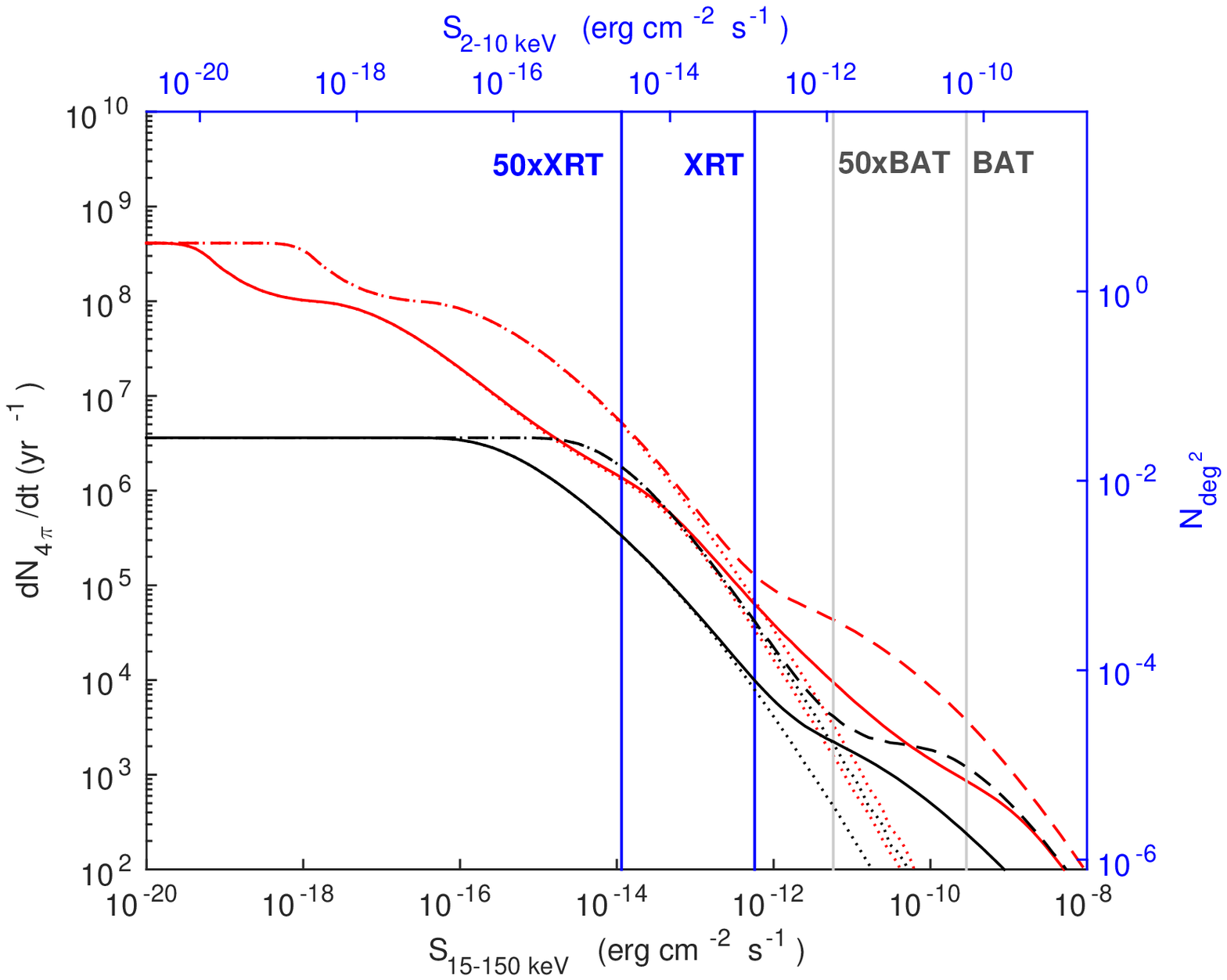}\includegraphics[width=3.4in]{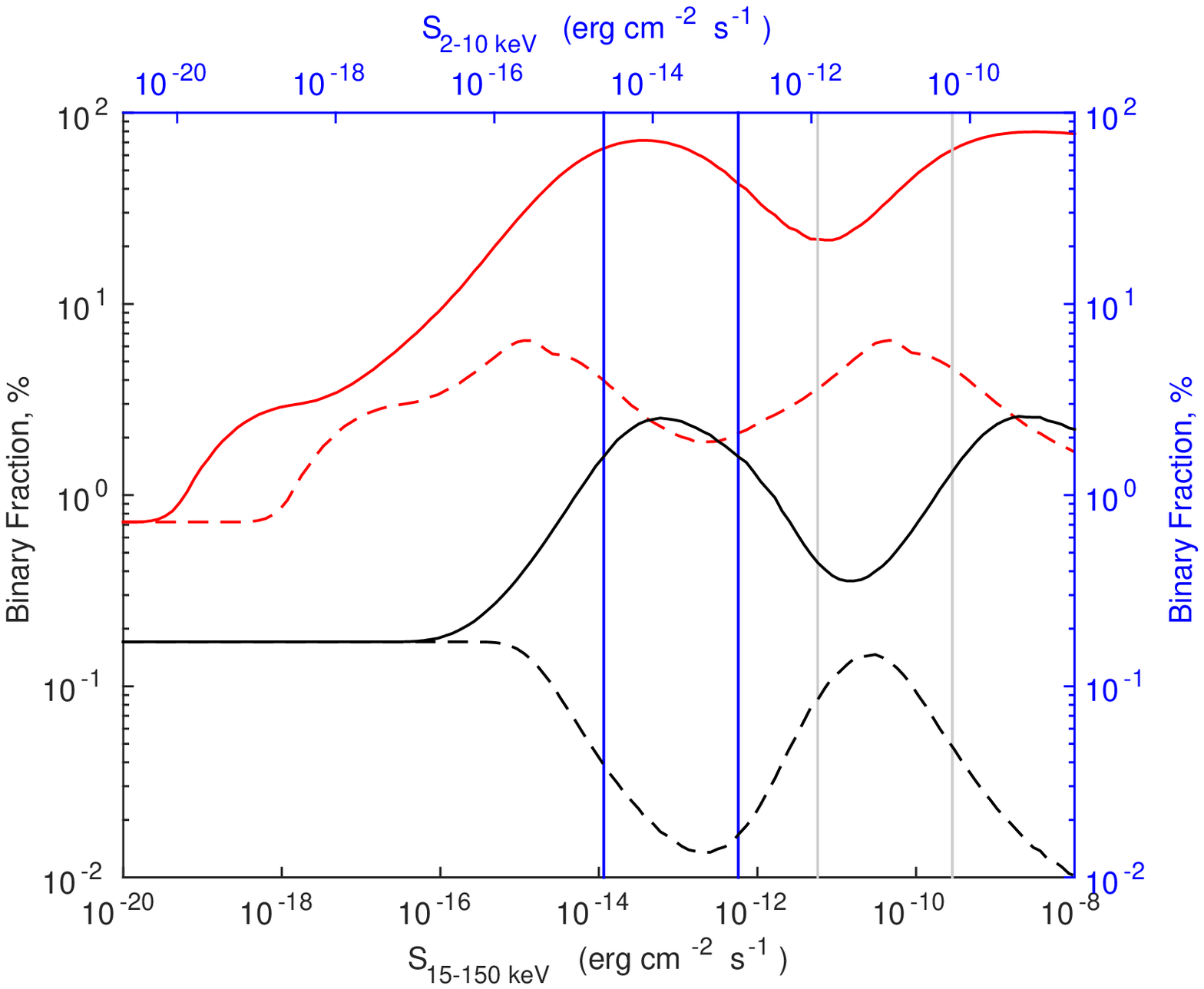}
\caption{Left: Cumulative all-sky number of  TDEs  brighter than the given flux limit $S_{15-150~\textrm{keV}}$. The bottom (black) axis labels refers to survey mode observations in the  15-150 keV band with the left vertical axis showing the TDE rates per sky per year. Grey vertical lines show 6-$\sigma$ BAT sensitivity of $2.88\times 10^{-10}$ erg cm$^{-2}$ s$^{-1}$ and a 50 times better  sensitivity of a  future instrument (``$50\times$BAT''). The upper (blue) coordinate system corresponds to snapshot mode observations in the 2-10 keV band with the right vertical axis showing the number of TDE observed per snapshot of integration time $t_{int}=1\times10^4$ s in a field of view of 1 deg$^2$ and blue vertical lines showing  6-$\sigma$ XRT sensitivity of $1.2\times 10^{-13}$ erg cm$^{-2}$ s$^{-1}$ and a 50 times better sensitivity of a instrument (``$50\times$XRT''). We show the results for SMBHs (black) and MBHs (red) for Models A (solid) and B (dashed). In all cases we have accounted for the contribution of both single and binary black holes, as well as for 10\% jetted events which are visible out to higher redshifts. The non-jetted contribution is shown with dotted lines in each case. Right: fraction of TDEs (in percent) above the flux limit which are contributed by binary systems. We use the same color code as on the left panel.}
\label{Fig:5}
\end{figure*}

Following \citet{Woods:1998},  the observed number of new events per year seen by BAT in the 15-150 keV band with peak flux larger than the flux limit $S_{lim}$ is given by 
\begin{equation}
\dot N_{TDE}^{S>S_{\lim}} = \int_0^{z_{max}} \int_{S_{15-150}>S_{lim}} \frac{ \dot N_{TDE}}{(1+z)}\frac{dV}{dz}dz dS,
\label{Eq:Surv}
\end{equation} 
where $S_{15-150}$ is the observed peak flux produced by each event.  This equation is appropriate for threshold experiments, such as BAT, observing  a population of transient sources that are standard candles in a peak flux. Figure  \ref{Fig:5} shows the total rates  of events with observed flux greater than $S_{lim}$ produced at all redshfits including jetted and non-jetted TDEs produced by both single and binary black holes. The black coordinate system  in the Figure shows the total number of hard X-ray events observed per year over the entire sky (field of view of $4\pi$) and for a 100\% duty cycle as a function of the telescope flux limit  $S_{lim}$, in the cases of our Model A and B and for $f_{occ} =0$ and 1.

As  seen from Figure \ref{Fig:5} where the contribution of non-jetted TDEs is labeled by a dotted line for each scenario, the expected number counts are dominated by jetted TDE  at the BAT sensitivity limit since the non-jetted contribution is negligible.  To compare our predictions to BAT observations we need to re-scale the rates correcting for the limited field of view and duty cycle of the telescope. First, assuming that three jetted TDEs were  detected  by BAT in 9 years of {\it Swift} lifetime with the duty cycle of 75\% over $4\pi /7$ of the sky we get $\dot N_{TDE} = 3$ yr$^{-1}$, while making use of the fact that the events were detected in three consecutive months (i.e., BAT sees 1 TDE per month) we get 112 TDEs per year. The latter number can be interpreted as a reasonable lower limit on the occurrence rate of jets and is just a factor of $\sim 2$ lower than our predictions for $f_{occ}=0$ (Model A) and a factor of $\sim 8$ for $f_{occ}=1$. The discrepancy could be explained by both observational limitations and modeling uncertainties, e.g.,  the assumed jetted fraction of 10\% might be overestimated. For a next-generation survey with 50 times better sensitivity than BAT, i.e. going from the BAT configuration to ``$50\times$BAT'',  our model predicts 11 times more sources for $f_{occ}=1$ and 3-9 more sources for $f_{occ}=0$ (see Table \ref{Table:2} for details).  

\begin{table}
		\begin{center}
			\begin{tabular}{llllllllll}
				\hline 
Model & Flux Limit &   $\frac{\dot N_{TDE}^{4\pi, A}}{10^3}$   &  $\frac{\dot N_{TDE}^{4\pi, A}}{10^3}$ &$\frac{\dot N_{TDE}^{4\pi, B}}{10^3}$ &   $\frac{\dot N_{TDE}^{4\pi, B}}{10^3}$ \\
				 &  & All  &  $z<3$ &All&   $z<3$     \\
		\hline   
	$f_{occ}=1$ & BAT   & $0.93 $  & $0.62$ & $4.3$ & $3.9$  \\		
			& $50\times$BAT  & $11$ & $9$ & $47$   &  $24$  \\		
		%	& Ideal  &$41\times 10^7$&  $3.6\times 10^7$   & $41\times 10^7$ & $3.6\times 10^7$  \\     
\hline
	$f_{occ}=0$ & BAT & $0.27 $   & $0.25 $   & $1.3 $  & $0.99 $   \\		
				&  $50\times$BAT & $2.4$  &  $1.5 $  & $4.6 $  & $3.6 $    \\		
		%		& Ideal &$3.6\times 10^6$  &  $1.8\times 10^6$ & $3.6\times 10^6$  &  $1.8\times 10^6$   \\				     
		      				\hline
	\end{tabular}
				\caption{For each model  we show the statistics of the observed events depending on the telescope sensitivity. TDE rates    per sky per year  (and divided by a factor of $10^3$, $\dot N_{TDE}^{4\pi, A}/10^3$) are shown for  Model A  for sources at all redshifts  (3rd column)  and at $0<z<3$ (4rd column); for  Model B ($\dot N_{TDE}^{4\pi, B}/10^3$) for all source redshifts  (5rd column) and $0<z<3$ (6rd column).}
				\label{Table:2}
\end{center}
\end{table}

In Table \ref{Table:3} we list the fraction of observable TDEs produced  by binaries  (in percents)  for each model and telescope sensitivity limit. At BAT sensitivity limit and in the case of high occupancy of IMBHs considerable fraction of observable TDEs are sourced by binary systems ($\sim 60\%$ for Model A and $\sim 5\%$ for Model B). As expected, because most of the faint systems are contributed by single MBHs, the fraction decreases as the sensitivity of the telescope improves. However, the decrease is non-monotonic as a function of $S_{lim}$ (as evident from the right panel of Figure \ref{Fig:5}), because of the contributions from  different  components (with/without jets, single and binary MBHs). In the case of a low occupancy, the contribution of binaries to the TDE sample is always below $2\%$. The contribution of binaries, and thus the occupation fraction of IMBHs, could be verified observationally by analyzing the variability of each event in and comparing to the models available in literature \citep{Liu:2009, Liu:2014, Coughlin:2016, Ricarte:2016}.

\begin{table}
		\begin{center}
			\begin{tabular}{llllllllll}
				\hline 
				 Model & Flux Limit &    $F_{Bin, A}^{BAT}$&    $F_{Bin, B}^{BAT}$     \\
		\hline   
				$f_{occ}=1$ & {\it Swift}   &  64\%  & 4.6\%   \\		
						& $50\times${\it Swift}  &  22\%  &   3.6\%   \\		
				            & Ideal  & 0.72\%  &  0.72\%  \\         	        			            
\hline
			$f_{occ}=0$ & {\it Swift} &  1.3\%  & 0.05\% \\		
						&  $50\times${\it Swift} &   0.5\%  & 0.09\%     \\		
				            & Ideal & 0.17\%  &  0.17\% \\				           				
				            \hline
	\end{tabular}
				\caption{For each model  we show the fraction of TDEs sourced by binaries at each flux limit   for Model A ($F_{Bin, A}^{BAT}$, column 3) and Model B ($F_{Bin, B}^{BAT}$, column 4).   }
				\label{Table:3}
\end{center}
\end{table}

Another observational mode is when the telescope takes a snapshot  of the same part of the sky  with a long exposure (integration time). The snapshot mode allows to probe a smaller portion of the sky with greater sensitivity than  is done in the survey mode. A telescope with integration  time $t_{int}$ will measure the following number of new events per frame \citep{Woods:1998}
\begin{equation}
N_{TDE}^{S>S_{\lim}} = f_{sky}\int_0^{z_{max}}\int_{S_{2-10}>S_{lim}} \frac{\dot N_{TDE}}{(1+z)}\frac{dV}{dz}dz dS
\label{Eq:Snap}
\end{equation} 
\begin{displaymath}
\times\min\left[t_{int}, t_{dur}(1+z)\right],
\end{displaymath}
where $t_{dur}$ is the time during which the event is above the sensitivity limit of the telescope,  and $f_{sky}$ is the sky fraction covered by the telescope. The typical integration time of a telescope such as XRT, $\sim 10^4$ seconds, is much  shorter than the typical duration of a TDE event ($\sim 10^6$ seconds, fall-back time), and thus,  $t_{dur}$ can be ignored compared to $t_{int}$. To complete our discussion of {\it Swift} capabilities in detecting TDEs we show in the blue coordinate system of Figure \ref{Fig:5}  the expected number of  events per typical exposure time of $10^4$ seconds in a field of view of 1 deg$^2$ assuming that all events shine at their peak luminosity. The number counts expected for the XRT sensitivity are  much smaller than unity (see Figure \ref{Fig:5} for details) meaning that XRT is good for follow-up missions but not to detect new TDEs, unless larger integration times are chosen. 

\begin{figure}
\centering
\includegraphics[width=3.4in]{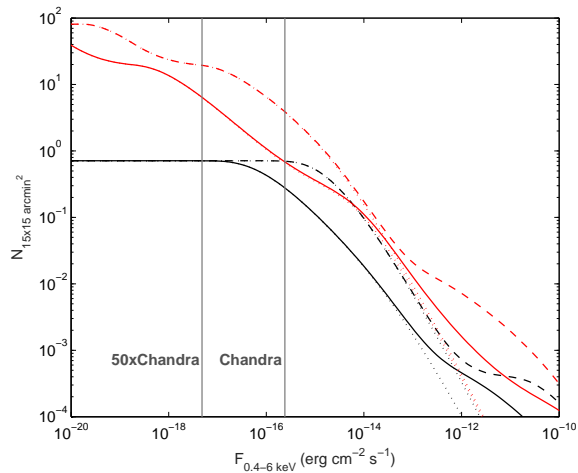}
\caption{Number counts per one deep exposure with a telescope such as the {\it Chandra X-ray Observatory}. The grey vertical lines show $6-\sigma$ sensitivity of $2.4\times 10^{-16}$ erg s$^{-1}$ and a 50 times better sensitivity, for an integration time of 4Ms.}
\label{Fig:6}
\end{figure}

The snapshot regime also applies to telescopes such as  {\it Chandra} which observe one  patch of the sky ($15\times15$ deg$^2$ in the case of {\it Chandra}) for a long time (more than $10^6$ seconds). The next generation upgrade of {\it Chandra}, called the {\it X-ray Surveyor}, is proposed   to have $\sim 30$ times bigger collecting area than {\it Chandra} and, therefore, better sensitivity \citep{Weisskopf:2015}. A small fraction of point sources in each  snapshot taken by {\it Chandra} (or the future {\it X-ray Surveyor}) could be TDEs and mistakenly identified as steady sources because of their long decay times.  To single them out, a succession of snapshots of the same field should be taken within a time interval longer than a year. Because of the very long integration time of  deep field survey, the TDE flux would decline below the flux limit in the course of the observation. To estimate the expected number of hidden TDEs in a {\it Chandra} deep field, we use Eq. (\ref{Eq:Snap}) and  assume that the light curve of each source fades according to  Eq. (\ref{Eq:Mfall}). Figure \ref{Fig:6} shows the increment in the observed number counts per one  snapshot as a telescope sensitivity limit improves. The number counts of SMBHs saturate at sensitivities $S_{lim} \sim 10^{-17}-10^{-16}$ erg s$^{-1}$, while the number counts of MBHs keep rising with decreasing $S_{lim}$. Comparing observed TDE luminosity function to the 6-$\sigma$ detection level ($\sim 2.4\times 10^{-16}$ erg cm$^{-2}$ s$^{-1}$ in $10^6$ s), we estimate number of TDEs in one {\it Chandra} deep field of $15\times 15$ arcmin$^2$ to be  $0.3-0.7$ for SMBHs and 0.7-4 for MBHs. For a future mission such as the {\it X-ray Surveyor} TDE number counts remain $\sim 0.7$ for SMBHs, while they evolve to  $\sim 6-20$ for MBHs. Therefore, non-detection of TDEs in a deep field would be a strong evidence for either a low occupation fraction of IMBHs, e.g., if they are kicked out of their parent halos as a result of mergers \citep{OLeary:2012} or a direct collapse scenario, e.g., works by \citet{Bromm:2003, Taeho:2016, Latif:2016, Chon:2016}.

Interestingly, current surveys with {\it Chandra} and {\it XMM-Newton}  find an exponential decline in the space density of luminous AGNs at $z > 3$ \citep{Brandt:2016} suggesting that MBHs might not exist at higher redshifts. As Table \ref{Table:2} shows, in our  Models A \& B  for current BAT sensitivity, 67\% \& 92\%   of all observable TDEs are expected originate at $z<3$ for MBHs (and 94\% \& 77\% for SMBHs); while for a 50 times more sensitive telescope, the corresponding numbers for MBHs are 83\% \& 52\% (66\% \& 78\% for SMBHs). For {\it Chandra} the corresponding numbers are 68\% \& 70\%  ($f_{occ}=1$) and  83\% \& 50\% ($f_{occ}=0$), while for its successor 52\% \& 35\% for MBHs and 49\% for SMBHs.

\begin{figure*}
\centering
\includegraphics[width=3.4in]{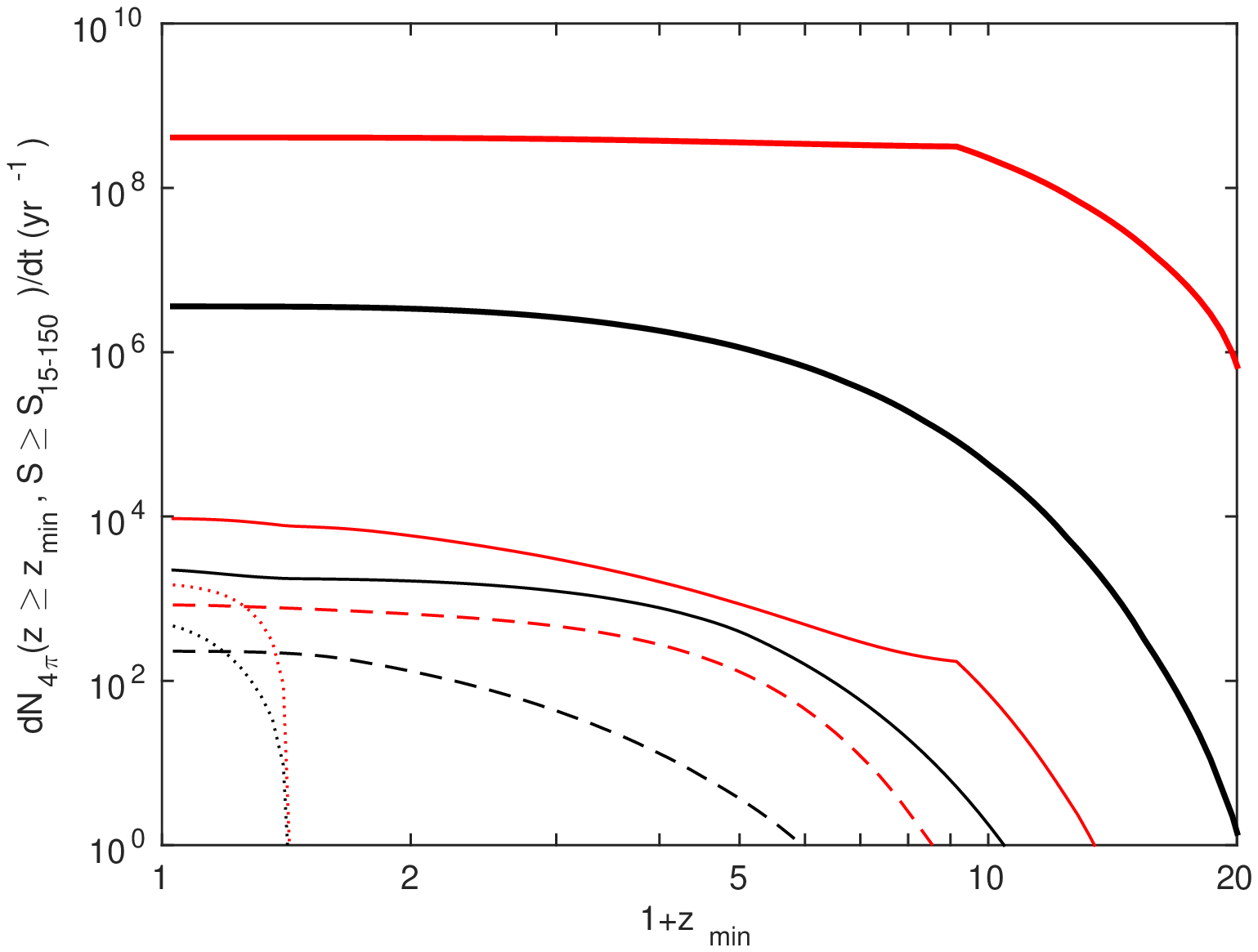}\includegraphics[width=3.4in]{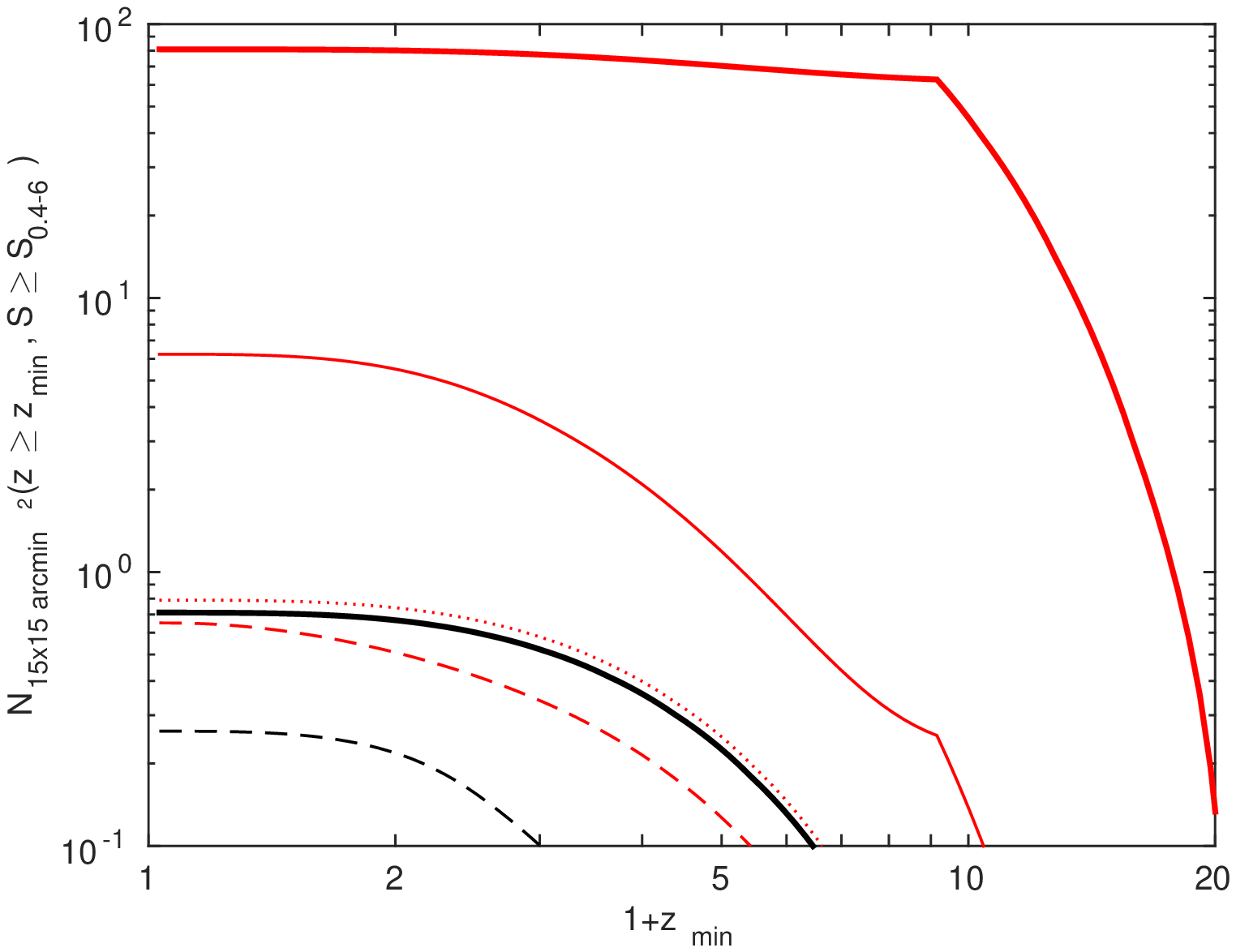}%\includegraphics[width=3.4in]{LF02_soft}%
\caption{TDE number counts originating from  $z\geq z_{min}$  for SMBHs (black) and MBHs (red) in Model A. In all cases we account for the contribution of both single and binary black holes, as well as for 10\% jetted events (which are visible out to higher redshifts). We also show a non-jetted population observed with a 50 times more sensitive telescope than present (dotted). Left: Hard X-ray counts observed in a survey mode over the whole sky per year for a telescope with sensitivity of $S_{lim}$ for BAT (dashed), future $50\times$BAT (thin solid) and an infinitely sensitive survey (thick solid). Right: Soft X-ray counts  observed in a snapshot mode over $15\times 15$ arcmin$^2$ field   for a telescope with sensitivity of for {\it Chandra} (dashed), $50\times${\it Chandra} (thin solid) and an ideal future survey which finds all sources (thick solid).}
\label{Fig:7}
\end{figure*}

The maximal redshift out to which TDEs can be detected depends on the telescope  sensitive. Figure \ref{Fig:7} shows the TDE rates in the survey mode (BAT, left) and number of TDEs observed per snapshot ({\it Chandra}, right) for several choices of $S_{lim}$ including present day instrument, a $50\times$ more sensitive telescope and an ideal detector which identifies  both jetted and non-jetted TDE accounting only for redshifts $z>z_{min}$. In other words, for each telescope sensitivity, we only account for the  events which originate at  $z_{min}$ or above. As we probe higher  redshifts, the expected number of sources drops because there are no sufficiently massive  halos to source sufficiently bright flares.  The left panel of Figure \ref{Fig:6} focuses on a BAT-like survey which is primarily sensitive to bright (mainly jetted) events.  It is evident that if TDEs do not source jets, prospects for observations with BAT would not be as bright, and TDEs would be observable only out to $z \sim 0.1$ with BAT and out to $z = 0.4$ with a 50 times more sensitive telescope than BAT. The redshifts at which number of observed TDE per year drops by a factor of 2 ($z_{50\%}^{BAT}$) and 10 ($z_{10\%}^{BAT}$) are listed in Table \ref{Table:4} for all models under consideration. Note that in some cases, $z_{50\%}^{BAT}$ does not change monotonically as a function of the sensitivity. This is because  the non-jetted events  become unobservable despite being more numerous while few jetted events are seen out to greater distances.  The right panel of Figure \ref{Fig:7} focuses on a {\it Chandra}-like instrument which has more sensitivity and integration time but a smaller field of view. If the IMBHs occupancy is high, 50\% of events observed by {\it Chandra} would originate from $z\lesssim 2$ (and $z\lesssim 2-4$  if observed by {\it Chandra} successors with the uncertainty arising from our luminosity modeling).  For completeness we also consider XRT. As expected, XRT is mainly sensitive to non-jetted TDEs and $50$\% of TDEs that can be observed by XRT are predicted to originate from $z\lesssim 0.4-1$. This is broadly consistent with the current sample  \citep{Komossa:2015} in which all  non-jetted TDEs are identified to be at $z\sim 0.405$ while the rare jetted events are observed at $z=0.353$, 0.89 and 1.186.

\begin{table}
		\begin{center}
			\begin{tabular}{llllllllll}
				\hline 
				 Model & Flux Limit &    $z_{50\%, A}^{BAT}$&    $z_{10\%, A}^{BAT}$&    $z_{50\%, B}^{BAT}$   & $z_{10\%, B}^{BAT}$      \\
		\hline   
				$f_{occ}=1$ & {\it Swift}  &  2.2&  4.4 & 1.2  &   2.6 \\		
						& $50\times${\it Swift}  &   1.3 &    3.8 & 2.8  &   5.3 \\		
				            & Ideal & 9.0 &  12.6 &9.0 &  12.6\\         	        			            
\hline
			$f_{occ}=0$ & {\it Swift} &    1.1 &  2.5 &2.0 &   3.2\\		
						&  $50\times${\it Swift} &   2.2 &    4.6 & 0.2  &   4.8\\		
				            & Ideal &  3.0 &  5.9 & 3.0 &    5.9\\				           				
				            \hline
	\end{tabular}
				\caption{The redshift $z_{min}$ so that 50\% (columns 3 and 5 for Model A and B respectively), and 10\% (columns 4 and 6) of observed BAT TDEs arrive from $z>z_{min}$. }
				\label{Table:4}
\end{center}
\end{table}

The signature of reionization, due to the evolving $M_{BH, min}$ in star-forming halos,  is evident in MBH cases and manifests itself as a cusp around  $z_{re}=8.8$. Because we assume instantaneous reionization, the feature is sharp. In a more realistic case of gradual reionization, the signature  is expected to show as a mild enhancement of the TDE rates at $z\gtrsim z_{re}$.

\section{Conclusions}
\label{Sec:sum}

Current observations pose only poor constraints on massive black hole growth at high redshifts as well as on the occupation fraction of IMBHs.  Flares from TDEs could reveal the population of otherwise dormant black holes allowing us to constrain the contribution of IMBHs. In this paper we have considered evolution of the observable TDE number counts  with X-ray telescopes including predictions for future missions. Our discussion of the black hole mass distribution included a model with $f_{occ}=1$ (all star forming halos are occupied by black holes) and $f_{occ}=0$ (only heavy halos host black holes of $M_{BH}>10^6 ~M_{\odot}$). These two scenarios provide an upper and lower limits for the expected number counts respectively. In addition, we considered two different prescriptions for the TDE luminosity: (i) Eddington luminosity, and (ii) luminosity proportional to the accretion rate. Even though current TDE observations suggest that the occupation fraction of IMBHs is very low with the majority of TDEs being produced by black holes of masses $\sim 10^6-10^8~M_\odot$, the results are far from being conclusive. Our study offers  new ways to constrain the occupation fraction of IMBHs at different cosmological redshifts, and our main conclusions are as follows
\begin{enumerate}
\item We show that jetted TDEs can be observed out to high redshifts and offer a unique probe of the occupancy of IMBHs. Earlier works have demonstrated that TDE rates  in merging systems are enhanced due to gravitational interactions of stars with binary black holes. Using this result we find that the higher is the occupation fraction of IMBHs the stronger is the impact of binaries on the total observed TDE rates. This is because with high IMBHs occupation there are enough progenitors to form binary systems. 

\item We show that TDEs sourced by binary black holes dominate the bright end of the X-ray luminosity function if the occupation fraction of IMBHs is high and if the TDE luminosity scales as Eddington. The shape of the TDE X-ray luminosity function is expected to show a unique signature of IMBHs in the form of two additional ``knees'', compared to the case with low IMBHs occupation. These features arise from the jetted and non-jetted contribution of black hole  binaries and are independent of our luminosity prescription (although the features are more evident when the TDE luminosity scales as the Eddington luminosity).  Therefore, for a complete TDE sample, the shape of the luminosity function  could be used to set an upper limit on the occupation fraction of IMBHs. Our results imply that, if $f_{occ} =1$ and the TDE luminosity scales as Eddington, the brightest events detected by BAT could be  associated with massive binary black holes. In this case the X-ray luminosity of TDE flares is expected to have excess of variability due to the binary interaction in addition to the typical power-law decay. 

\item   The fraction of observable TDE that are generated by binaries depends on the luminosity prescription as well as on the sensitivity of the telescope. With current X-ray telescopes,  we expect to see $>2\%$ and up to 64\% of TDEs produced by binary black holes if the occupation fraction of IMBHs is high; while the fraction is at most 1.3\% if the occupation fraction of IMBHs is low. Since dimmer events are mainly contributed by single black holes, the binary fraction drops with the telescope sensitivity. 

\item Detection of TDEs in deep field observations by {\it Chandra} and future missions  would provide a smoking gun signature of IMBHs. We find that in the case when only SMBHs contribute, TDEs are not expected in   {\it Chandra} deep fields; while if the IMBHs  occupation fraction is high, some point sources in the archival data of X-ray deep field surveys may be TDEs. To identify such events one should compare two images of the same deep field separated by an interval of at least a year.  Non-detection of TDEs from high redshifts can set upper limits on the occupation fraction of IMBHs and constrain direct collapse scenarios of SMBH formation, e.g., works by \citet{Bromm:2003, Taeho:2016, Latif:2016, Chon:2016}. 

\item  Increasing sensitivity of X-ray telescopes  by a factor of 50 comparing to current instruments will increase the expected number counts by a factor of  $4- 10$ for a BAT-like mission and a factor of $20-40$ for an XRT-like mission with 1ks itegration time.  For a deep field survey the improvement strongly depends on the occupation fraction of IMBHs. Current sensitivity is enough to resolve most TDEs if $f_{occ}=0$, and, therefore, improvement in sensitivity would not yield new events in this case. However, if $f_{occ}=1$, improving the sensitivity by a factor of 50 would increase the number of TDEs per snapshot by a factor of 5-10.    

\end{enumerate}

Comparing our model to existing observations,  a low occupation fraction is suggested (see also Stone \& Metzger 2016). However, observations are far from being conclusive and  it is still unclear why TDEs sourced by  IMBHs with masses below $10^6~M_\odot$ are not observed.  Several possible explanations can follow: IMBHs are kicked out of their dark matter halos as a result of mergers \citep{OLeary:2012}, SMBHs are formed from massive seeds in massive halos, observational selection effects exclude TDEs around IMBHs, the assumption of an isothermal density distribution is less suitable  for smaller galaxies, or low mass systems are more sensitive to AGN feedback which expels  gas from halo and limits star formation leading to inefficient replenishing of the loss cone. 

If $f_{occ}$ is high at low black hole masses and IMBH binaries indeed play a role in sourcing TDEs, these binaries would also produce gravitational waves on their approach to  coalescence. eLISA should be sensitive to MBH binaries over a wide range of total masses and mass ratios, e.g., systems with total mass  $ \gtrsim 10^5$ M$_\odot$ and mass ratios of $\gtrsim 0.1$ will yield  signal to noise ratio of $> 20$ out to $z = 4$ \citep{Amaro:2012}. Therefore, in the future one could use a cross-correlation between the  stochastic gravitational wave background and the spatial distribution of brightest TDEs to constrain the role of IMBHs in TDE production.  If the two quantities correlate, IMBHs must make a significant contribution to TDEs. 

In addition to the X-ray observations discussed here, jetted TDEs may also be bright in the radio band \citep{Zauderer:2011, Levan:2011}. Snapshot rates of jetted TDEs in radio band have been computed by \citet{vanVelzen:2011}, and prospects for the detection of jetted TDEs by the Square Killometer Array out to $z\sim 2$ as well as the synergy between radio and X-ray observations was discussed \citep{Donnarumma:2015b, Donnarumma:2015, Rossi:2015}.  At present there is no consensus as to whether  X-ray emission by jetted AGN should correlate with their radio emission. In particular, the peak emissivity of Sw1644+57 appeared 100 days after the BAT trigger and the Lorentz factor of radio jet was found to be $\Gamma \sim 2$, much lower than what was observed in X-rays right after  detection  ($\Gamma \sim 10$). The radio data require a different energy injection mechanism such as an outflow with a distribution of Lorentz factors \citep{Berger:2012, Generozov:2016}.  Therefore, we did not use our simplex model to predict radio emission from TDEs. 

Tidal disruptions occurring at high redshifts can reveal the seeds of quasars. In this paper we have assumed that the high-redshift population resembles that of today; however, its properties might evolve with redshift. In particular, simulations show that first stars were much more massive (up to $10^3~M_\odot$) than present-day stars and could serve as an additional population of seeds. Star formation in high-redshift, low-mass halos strongly depends on feedback processes such as photoheating feedback, as well as AGN and supernovae feedback. The evolution of the TDE number counts  with redshift could serve as a smoking gun for these processes.  
 
\section{Acknowledgments}
We thank J. Guillochon and N. Stone for helpful comments on the manuscript. We also thank A. Sadowski, Y.-F. Jiang, and A. Siemiginowska for useful discussions. This work was supported in part by Harvard's Black Hole Initiative, which is supported by a grant from the John Templeton Foundation.


\begin{thebibliography}{14}

\bibitem[\protect\citeauthoryear{Adam et al.}{2016}]{Planck:2016}
Adam, R., et al., 2016, arXiv:1605.03507

\bibitem[\protect\citeauthoryear{Amaro-Seoane et al.}{2012}]{Amaro:2012}
 Amaro-Seoane, P., Aoudia, S., Babak, S., Binétruy, P., Berti, E., et al., 2012, CQG 2914016 
 
 \bibitem[\protect\citeauthoryear{Arcavi}{2014}]{Arcavi:2014}
 Arcavi I., et al., 2014, ApJ, 793, 38
 
\bibitem[\protect\citeauthoryear{Bade et al.}{1996}]{Bade:1996}
 Bade, N., et al., 1996, A\&A 309, L35



\bibitem[\protect\citeauthoryear{Baldassare et al.}{2015}]{BALDASSARE:2015}
Baldassare, V. F., Reines, A. E., Gallo, E., Greene, J. E., 2015, ApJ, 809, 14

\bibitem[\protect\citeauthoryear{Baldassare et al.}{2016}]{BALDASSARE:2016}
Baldassare, V. F., Reines, A. E., Gallo, E., Greene, J. E., arXiv:1609.07148

\bibitem[\protect\citeauthoryear{Ba{\~n}ados et al.}{2015}]{Banados:2015}
Ba{\~n}ados, E., Venemans, B. P., Morganson, E., Hodge, J., Decarli, R., et al., 2015, ApJ, 804, 118

\bibitem[\protect\citeauthoryear{Bar-Or \& Alexander}{2016}]{Bar-Or:2016}
Bar-Or B., Alexander T., 2016, ApJ, 820, 129

\bibitem[\protect\citeauthoryear{Barkana}{2016}]{Barkana:2016}
Barkana, R., 2016, PhR, 645, 1


\bibitem[\protect\citeauthoryear{Barthelmy et al.}{2005}]{Barthelmy:2005}
Barthelmy, S. D., Barbier, L. M., Cummings, J. R., Fenimore, E. E., Gehrels, N., et al., 2005, SSR, 120, 143 

\bibitem[\protect\citeauthoryear{Berger et al.}{2012}]{Berger:2012}
Berger, E., Zauderer, A., Pooley, G., Soderberg, A. M., Sari, R., et al.,  2012, ApJ, 748, 36


\bibitem[\protect\citeauthoryear{Bloom et al.}{2011}]{Bloom:2011}
Bloom, J. S., Giannios, D., Metzger, B. D., et al.,  2011, Science, 33, 203

\bibitem[\protect\citeauthoryear{Bower et al.}{2013}]{Bower:2013}
Bower, G. C., Metzger, B. D., Cenko, S. B., Silverman, J. M., Bloom,
J. S., 2013, ApJ, 763, 84


\bibitem[\protect\citeauthoryear{Brandt \& Vito}{2016}]{Brandt:2016}
Brandt, W. N., Vito, F.,arXiv:1609.07527

\bibitem[\protect\citeauthoryear{Bromm \& Loeb}{2003}]{Bromm:2003}
Bromm, V., Loeb, A.,  2003, ApJ, 596, 34
 
\bibitem[\protect\citeauthoryear{Brown et al.}{2015}]{Brown:2015}
Brown, G. C., Levan, A. J., Stanway, E. R., Tanvir, N. R., Cenko, S. B., et al., 2015, MNRAS, 452, 4297


\bibitem[\protect\citeauthoryear{Burrows et al.}{2011}]{Burrows:2011}
Burrows, D. N., Kennea, J. A., Ghisellini, G., et al.,  2011, Nature, 476, 421 

\bibitem[\protect\citeauthoryear{Cappelluti et al.}{2009}]{Cappelluti:2009}
Cappelluti, N.,  Ajello, M.,  Rebusco, P.,  Komossa, S., Bongiorno, A., et al., 2009, A\&A, 495, 9

\bibitem[\protect\citeauthoryear{Carter \& Luminet}{1983}]{Carter:1983}
Carter, B., Luminet, J.-P., 1983, AAP, 121, 97

\bibitem[\protect\citeauthoryear{Cenko et al.}{2012}]{Cenko:2012}
Cenko, S. B., et al. 2012, ApJ, 753, 77

\bibitem[\protect\citeauthoryear{Chen et al.}{2009}]{Chen:2009}
Chen, X., Madau, P., Sesana, A., Liu, F. K., 2009, ApJ, 697, 149

\bibitem[\protect\citeauthoryear{Chen et al.}{2011}]{Chen:2011}
Chen, X., Sesana, A.,  Madau, P., Liu, F. K., 2011, ApJ, 729, 13

\bibitem[\protect\citeauthoryear{Chon et al.}{2016}]{Chon:2016}
Chon, S., Hirano, S., Hosokawa, T., Yoshida, N., arXiv:160308923

\bibitem[\protect\citeauthoryear{Cohen et al.}{2016}]{Cohen:2016}
Cohen, A., Fialkov, A., Barkana, R., 2016, MNRAS, 459, 90

\bibitem[\protect\citeauthoryear{Chornock et al.}{2014}]{Chornock:2014}
Chornock, R., et al., 2014, ApJ, 780, 44

\bibitem[\protect\citeauthoryear{Cohn \& Kulsrud}{1978}]{Cohn:1978}
Cohn, H., Kulsrud, R. M., 1978, ApJ, 226, 1087

\bibitem[\protect\citeauthoryear{Colpi}{2014}]{Colpi:2014}
 Colpi, M., 2014, SSRv, 183, 189
 
\bibitem[\protect\citeauthoryear{Coughlin et al.}{2016}]{Coughlin:2016}
Coughlin, E. R., Armitage, P. J., Nixon, C., Begelman, M. C., arXiv:1608.05711

\bibitem[\protect\citeauthoryear{Crumley et al.}{2016}]{Crumley:2016}
Crumley, P., Lu, W., Santana, R., Hernandez, R. A., Kumar, P., Markoff, S., 2016, MNRAS, 460, 396

\bibitem[\protect\citeauthoryear{Dai et al.}{2015}]{Dai:2015}
Dai, L., McKinney, J. C., Miller, M. C., 2015, ApJ, 812, L39


\bibitem[\protect\citeauthoryear{Donley et al.}{2002}]{Donley:2002}
Donley, J. L., et al., 2002, AJ, 124, 1308 

\bibitem[\protect\citeauthoryear{Donnarumma et al.}{2015}]{Donnarumma:2015}
Donnarumma, I., Rossi, E. M., Fender, R., Komossa, S., Paragi, Z., et al., arXiv:1501.04640


\bibitem[\protect\citeauthoryear{Donnarumma \& Rossi}{2015}]{Donnarumma:2015b}
Donnarumma, I., Rossi, E. M., 2015, ApJ, 2015, 803, 36 

\bibitem[\protect\citeauthoryear{Esquej et al.}{2008}]{Esquej:2008}
Esquej, P., et al., 2008, A\&A, 489, 543



\bibitem[\protect\citeauthoryear{Evans \& Kochaneck}{1989}]{Evans:1989}
 Evans, C. R., Kochaneck, C. S., 1989, ApJ, 346, L13
 
\bibitem[\protect\citeauthoryear{Fakhouri et al.}{2010}]{Fakhouri:2010}
Fakhouri, O., Ma, C.-P., Boylan-Kolchin, M., 2010, MNRAS, 406, 2267

\bibitem[\protect\citeauthoryear{Frank \& Rees}{1976}]{Frank:1976}
Frank, J., Rees, M. J., 1976, MNRAS, 176, 633

\bibitem[\protect\citeauthoryear{Gezari et al.}{2006}]{Gezari:2006}
Gezari S., Martin D. C., Milliard B., Basa S., Halpern J. P., et al., 2006, ApJL, 653, L25


\bibitem[\protect\citeauthoryear{Gezari et al.}{2008}]{Gezari:2008}
Gezari, S., et al,. 2008, ApJ, 676, 944 

\bibitem[\protect\citeauthoryear{Gezari et al.}{2009}]{Gezari:2009}
Gezari, S., et al., ApJ., 2009, 698, 1367

\bibitem[\protect\citeauthoryear{Gezari et al.}{2012}]{Gezari:2012}
 Gezari, S., et al., 2012, Nature, 485, 217



\bibitem[\protect\citeauthoryear{Generozov et al.}{2016}]{Generozov:2016}
Generozov, A., Mimica, P., Metzger, B. D., Stone, N. C., Giannios, D., Aloy, M. A.,  arXiv:1605.08437

\bibitem[\protect\citeauthoryear{Ghisellini et al.}{2010}]{Ghisellini:2010}	
Ghisellini, G., Della Ceca, R., Volonteri, M., Ghirlanda, G., Tavecchio, F., 2010, MNRAs, 405, 387 

\bibitem[\protect\citeauthoryear{Graham}{2015}]{Graham:2015}
Graham, A., arXiv:1501.02937


\bibitem[\protect\citeauthoryear{Granot \& Sari}{2002}]{Granot:2002}
Granot, J.,  Sari, R., 2002, ApJ, 568, 820

\bibitem[\protect\citeauthoryear{Greene \& Ho}{2007}]{Greene:2007}
Greene, J. E., Ho, L. C., 2007, ApJ, 667, 131

\bibitem[\protect\citeauthoryear{Greene}{2012}]{Greene:2012}
Greene, J. E.,  2012, NatCo, 3, 1304


\bibitem[\protect\citeauthoryear{Guillochon \& Ramirez-Ruiz}{2013}]{Guillochon:2013}
Guillochon, J.,  Ramirez-Ruiz, E., 2013, ApJ, 767, 25


\bibitem[\protect\citeauthoryear{Guillochon et al.}{2015}]{Guillochon:2015}
Guillochon, J., McCourt, M., Chen, X., Johnson, M. D., Berger, E., 2015 ://arxiv.org/abs/1509.08916

\bibitem[\protect\citeauthoryear{Guillochon \& Loeb}{2015b}]{Guillochon:2015b}
Guillochon, J., Loeb, A., 2015, ApJ, 806, 124

\bibitem[\protect\citeauthoryear{Guillochon \& Ramirez-Ruiz}{2015c}]{Guillochon:2015c}
Guillochon, J., Ramirez-Ruiz, E., 2015, ApJ, 809, 166

\bibitem[\protect\citeauthoryear{Guillochon \& McCourt}{2016}]{Guillochon:2016}
Guillochon, J., McCourt, M., arXiv: 1609.08160

\bibitem[\protect\citeauthoryear{Halpern et al.}{2004}]{Halpern:2004}
Halpern, J. P.,  Gezari, S.,  Komossa, S., 2004, ApJ, 604, 572


\bibitem[\protect\citeauthoryear{Hayasaki et al.}{2016}]{Hayasaki:2016}
Hayasaki, K., Stone, N., Loeb, A., 2016, MNRAS, 461, 3760

\bibitem[\protect\citeauthoryear{Hill}{1975}]{Hill:1975}
Hill, J. G., 1975, Nature, 254, 295


\bibitem[\protect\citeauthoryear{Holoien et al.}{2014}]{Holoien:2014}
Holoien, T. W.-S., Prieto, J. L., Bersier, D., Kochanek, C. S., Stanek, et al., 2014, MNRAS, 445, 3263


\bibitem[\protect\citeauthoryear{Hopman \& Alexander}{2006}]{Hopman:2006}
Hopman, C., Alexander, T., 2006, ApJ, 645, 1152

\bibitem[\protect\citeauthoryear{Inayoshi et al.}{2016}]{Inayoshi:2016}	
Inayoshi, K., Haiman, Z., Ostriker, J. P., 2016, MNRAS, 459, 3738

\bibitem[\protect\citeauthoryear{Ivanov et al.}{2005}]{Ivanov:2005}
Ivanov, P. B., Polnarev, A. G., Saha, P., 2005, MNRAS, 358, 1361

\bibitem[\protect\citeauthoryear{Farrell et al.}{2009}]{Farrell:2009}
Farrell, S, A., Webb, N. A., Barret, D., Godet, O., Rodrigues, J. M., 2009, Nature, 460, 73


\bibitem[\protect\citeauthoryear{Jiang et al.}{2007}]{Jiang:2007}
Jiang, L., Fan, X., Ivezi{\' c}, {\u Z}., Richards, G. T., Schneider, D. P., et al., 2007,ApJ, 656, 680

\bibitem[\protect\citeauthoryear{Jiang et al.}{2014}]{Jiang:2014}
Jiang, Y.-F., Stone, J. M., Davis, S. W., 2014, ApJ, 796, 106

\bibitem[\protect\citeauthoryear{Kara et al.}{2016}]{Kara:2016}
Kara, E., Miller, J. M., Reynolds, C., Dai, L., 2016, Nature, 535, 388

\bibitem[\protect\citeauthoryear{Kawamuro et al.}{2016}]{Kawamuro:2016}
Kawamuro, T., Ueda, Y., Shidatsu, M., Hori, T., Kawai, N., et al., 2016, PASJ, 68, 58 


\bibitem[\protect\citeauthoryear{Kesden}{2012}]{Kesden:2012}
Kesden, M., 2012, PRD, 85, 4037

\bibitem[\protect\citeauthoryear{Khabibullin \& Sazonov}{2014}]{Khabibullin:2014}
Khabibullin, I., Sazonov, S., 2014, MNRAS, 444, 1041


\bibitem[\protect\citeauthoryear{Kochanek}{2016}]{Kochanek:2016}
Kochaneck,  C. S., arXiv:1601.06787


\bibitem[\protect\citeauthoryear{Komossa}{2015}]{Komossa:2015}
Komossa, S., 2015, JHEA, 7, 148

\bibitem[\protect\citeauthoryear{Komossa \& Bade}{1999}]{Komossa:1999}
Komossa, S., Bade, N., 1999, A\&A, 343, 775


\bibitem[\protect\citeauthoryear{Kormendy \& Ho}{2013}]{Kormendy:2013}
Kormendy, J., Ho, L. C., 2013, ARA\&A, 51, 511


\bibitem[\protect\citeauthoryear{Lacy}{1982}]{Lacy:1982}
Lacy, J. H., Townes, C. H.,  Hollenbach, D. J., 1982, ApJ, 262, 120 

\bibitem[\protect\citeauthoryear{Latif \& Ferrara}{2016}]{Latif:2016}
Latif, M., Ferrara, A., arXiv:160507391

\bibitem[\protect\citeauthoryear{Lemons et al.}{2015}]{Lemons:2015}
Lemons, S. M., Reines, A. E., Plotkin, R. M., Gallo, E., Greene, J. E., 2015, ApJ, 805, 12


\bibitem[\protect\citeauthoryear{Lezhnin \& Vasiliev}{2015}]{Lezhnin:2015}
Lezhnin, K., Vasiliev, E., 2015, ApJL, 808, L5

\bibitem[\protect\citeauthoryear{Lezhnin \& Vasiliev}{2016}]{Lezhnin:2016}
Lezhnin, K., Vasiliev, arXiv:1609.00009


\bibitem[\protect\citeauthoryear{Levan et al.}{2011}]{Levan:2011}
Levan, A. J., Tanvir, N. R., Cenko, S. B., 2011, Science, 33, 199

\bibitem[\protect\citeauthoryear{Li et al.}{2015}]{Li:2015}	
Li, S., Liu, F. K., Berczik, P., Spurzem, R., arXiv:1509.00158


\bibitem[\protect\citeauthoryear{Li et al.}{2015b}]{LiNaoz:2015}	
Li, G., Naoz, S., Kocsis, B., Loeb, A., 2015b, MNRAS, 451, 1341


\bibitem[\protect\citeauthoryear{Liu et al.}{2009}]{Liu:2009}	
Liu, F. K., Li, S., Chen, X., 2009, ApJ, 706, L133

\bibitem[\protect\citeauthoryear{Liu \& Chen}{2013}]{Liu:2013}	
Liu, F. K., Chen, X., 2013, ApJ, 767, 18

\bibitem[\protect\citeauthoryear{Liu et al.}{2014}]{Liu:2014}	
Liu, F. K., Li, S., Komossa, S., 2014, ApJ, 786, 103

\bibitem[\protect\citeauthoryear{Lightman \& Shapiro}{1977}]{Lightman:1977}	
Lightman, A. P., Shapiro, S. L., 1977, ApJ, 211 244

\bibitem[\protect\citeauthoryear{Loeb \& Furlanetto}{2013}]{LoebFurlanetto:2013}	
Loeb, A., Furlanetto, S., 2013,  The First Galaxies in the
Universe, Princeton University Press (Princeton)

\bibitem[\protect\citeauthoryear{Luo et al.}{2008}]{Luo:2008}	
Luo, B., et al. 2008, ApJ, 674, 122

\bibitem[\protect\citeauthoryear{Machacek et al.}{2001}]{Machacek:2001}	
Machacek, M. E., Bryan, G. L., Abel, T., 2001, ApJ, 548, 509

\bibitem[\protect\citeauthoryear{Magorrian  \& Tremaine}{1999}]{Magorrian:1999}	
Magorrian, J., Tremaine, S.,  1999, MNRAS, 309, 447

\bibitem[\protect\citeauthoryear{Maksym et al.}{2010}]{Maksym:2010}	
Maksym, W. P., et al., 2010, ApJ, 722, 1035


\bibitem[\protect\citeauthoryear{McConnell \& Ma}{2013}]{McConnell:2013}	
McConnell, N. J., Ma, C.-P., 2013, ApJ, 764, 184

\bibitem[\protect\citeauthoryear{McKinney et al.}{2014}]{McKinney:2014}
McKinney, J. C., Tchekhovskoy, A. Sadowski, A., Narayan, R., 2014, MNRAS 441, 3177

\bibitem[\protect\citeauthoryear{McKinney et al.}{2015}]{McKinney:2015}	
McKinney, J. C., Dai, L., Avara, M. J., 2015, MNRAS, 454, 6

\bibitem[\protect\citeauthoryear{Merritt}{2015}]{Merritt:2015}	
Merritt, D., 2015, ApJ, 804, 128

\bibitem[\protect\citeauthoryear{Merritt \& Poon}{2004}]{Merritt:2004}	
Merritt, D., Poon M.-Y., 2004, ApJ, 606, 788

\bibitem[\protect\citeauthoryear{Merritt \& Wang}{2005}]{Merritt:2005}	
Merritt, D., Wang J., 2005, ApJL, 621, L101

\bibitem[\protect\citeauthoryear{Metzger et al.}{2012}]{Metzger:2011} 
Metzger, B. D., Giannios, D., Mimica, P., 2012, MNRAS, 420, 3528

\bibitem[\protect\citeauthoryear{Miller et al.}{2015}]{Miller:2015} 
Miller, B. P., Gallo, E., Greene, J. E., Kelly, B. C., Treu, T., et al., 2015, ApJ, 99, 98

\bibitem[\protect\citeauthoryear{Mimica et al.}{2015}]{Mimica:2015} 
Mimica, P., Giannios, D., Metzger, B. D., Aloy, M. A., 2015, MNRAS,
450, 2824

\bibitem[\protect\citeauthoryear{Moran et al.}{2014}]{Moran:2014}
Moran, E. C., Shahinyan, K., Sugarman, H. R., Velez, D. O., Eracleous, M., 2014, AJ, 148, 136


\bibitem[\protect\citeauthoryear{O'Leary \& Loeb}{2012}]{OLeary:2012}
O'Leary, R. M., Loeb, A., 2012, MNRAS, 421, 2737


\bibitem[\protect\citeauthoryear{Pasham et al.}{2015}]{Pasham:2015}
Pasham D. R., Cenko S. B., Levan A. J., Bower G. C., Horesh A., et al., 2015, ApJ, 805, 68


\bibitem[\protect\citeauthoryear{Perets et al.}{2006}]{Perets:2006}	
Perets, H., Hopman, C., Alexander, T., 2006, ApJ, 656, 709

\bibitem[\protect\citeauthoryear{Piran et al.}{2015}]{Piran:2015}	
Piran, T., Sądowski, A., Tchekhovskoy, A., 2015, MNRAS, 453, 157 

\bibitem[\protect\citeauthoryear{Phinney}{1989}]{Phinney:1989}
Phinney, E. S., 1989, Nature, 340, 595

\bibitem[\protect\citeauthoryear{Rauch \& Tremaine}{1996}]{Rauch:1996}
Rauch, K., Tremaine, S., 1996, New Astron., 1, 149


\bibitem[\protect\citeauthoryear{Rees}{1988}]{Rees:1988}	
Rees, M. J., 1988, Nature, 333, 523

\bibitem[\protect\citeauthoryear{Rees}{1990}]{Rees:1990}	
Rees, M. J., 1990, Science, 247, 817

\bibitem[\protect\citeauthoryear{Reines}{2013}]{Reines:2013}	
Reines, A. E., Greene, J. E.,  Geha, M. 2013, ApJ, 775, 116

\bibitem[\protect\citeauthoryear{Ricarte et al.}{2016}]{Ricarte:2016}	
Ricarte, A., Natarajan, P., Dai, L., Coppi, P., 2016, MNRAS, 458, 1712

\bibitem[\protect\citeauthoryear{Rossi et al.}{2015}]{Rossi:2015}	
Rossi, E. M., Donnarumma, I., Fender, R., Jonker, P., Komossa, S., et al., arXiv:1501.02774

\bibitem[\protect\citeauthoryear{Roth et al.}{2016}]{Roth:2016}	
Roth, N.,  Kasen, D., Guillochon, J., Ramirez-Ruiz, E., 2016, ApJ, 827, 3


\bibitem[\protect\citeauthoryear{Ryu et al.}{2016}]{Taeho:2016}
Ryu, T., Tanaka, T. L., Perna, R., Haiman, Z., 2016, MNRAS, 460, 4122

\bibitem[\protect\citeauthoryear{Sadowski et al.}{2015a}]{Sadowski:2015a}
Sądowski, A., Narayan, R., Tchekhovskoy, A., Abarca, D., Zhu, Y., McKinney, J. C., 2015, MNRAS, 447, 49

\bibitem[\protect\citeauthoryear{Sadowski \& Narayan}{2015b}]{Sadowski:2015b}
Sądowski, A., Narayan, R., 2015, MNRAS, 453, 3213

\bibitem[\protect\citeauthoryear{Saglia et al.}{2016}]{Saglia:2016}
Saglia, R. P. et al., 2016, ApJ, 818, 47 


\bibitem[\protect\citeauthoryear{Sakurai et al.}{2016}]{Sakurai:2016}	
Sakurai, Y., Inayoshi, K., Haiman, Z., 2016, MNRAS, 461, 4496

\bibitem[\protect\citeauthoryear{Saxton et al.}{2012}]{Saxton:2012}	
Saxton R. D., Read A. M., Esquej P., Komossa S., Dougherty S., et al.,  2012, A\&A, 541, 106

\bibitem[\protect\citeauthoryear{Saxton et al.}{2016}]{Saxton:2016}	
Saxton R. D., Read A. M., Komossa S., Lira, P.,  Alexander, K. D., Wieringa, M. H., arXiv:1610.01788


\bibitem[\protect\citeauthoryear{Sheth \& Tormen}{1999}]{Sheth:1999}
Sheth, R. K.,  Tormen, G., 1999, MNRAS, 308, 119

\bibitem[\protect\citeauthoryear{Shiokawa et al.}{2015}]{Shiokawa:2015}
Shiokawa, H., Krolik, J. H., Cheng, R. M., Piran, T., Noble, S. C., 2015, ApJ, 804, 85

\bibitem[\protect\citeauthoryear{Sobacchi \& Mesinger}{2013}]{Bacchic:2013}
Sobacchi, E., Mesinger, A., 2013, MNRAS, 432, 3340

\bibitem[\protect\citeauthoryear{Stern et al.}{2004}]{Stern:2004}
Stern D., van Dokkum P. G., Nugent P., Sand D. J., Ellis R. S., et al., 2004, ApJ, 612, 690

\bibitem[\protect\citeauthoryear{Stone et al.}{2013}]{Stone:2013}
Stone, C. N., Sari, R., Loeb, A., 2013, MNRAS, 435, 1809


\bibitem[\protect\citeauthoryear{Stone \& Metzger}{2016}]{Stone:2016}
Stone, C. N., Metzger, B. D., 2016, MNRAS, 455, 859

\bibitem[\protect\citeauthoryear{Stone et al.}{2016}]{Stone:2016KO}
Stone, C. N., Kuepper, A. H. W., Ostriker, J. P., arXiv:1606.01909

\bibitem[\protect\citeauthoryear{Tegmark et al.}{1997}]{Tegmark:1997}
Tegmark, M., Silk, J., Rees, M. J., Blanchard, A., Abel, T., Palla, F., 1997, ApJ, 474, 1

\bibitem[\protect\citeauthoryear{Tchekhovskoy et al.}{2014}]{Tchekhovskoy:2014}
Tchekhovskoy, A., Metzger, B. D., Giannios, D., Kelley, L. Z., 2014, MNRAS, 437, 2744

	
	
\bibitem[\protect\citeauthoryear{Thomas et al.}{2016}]{Thomas:2016}
Thomas, J., Ma, C.-P., McConnell, N. J., Greene, J. E., Blakeslee, J. P., Janish, R., 2016, Nature, 532, 340


\bibitem[\protect\citeauthoryear{van Velzen et al.}{2011}]{vanVelzen:2011}
van Velzen, S.,  Farrar, G. R., Gezari, S., et al., 2011, ApJ, 741, 73

\bibitem[\protect\citeauthoryear{van Velzen et al.}{2013}]{vanVelzen:2013}
van Velzen, S., Frail, D. A., K{\"o}rding, E., Falcke, H., 2013, A\&A,
552, A5


\bibitem[\protect\citeauthoryear{van Velzen \& Farrar}{2014}]{vanVelzen:2014}
van Velzen, S.,  Farrar, G. R., 2014, ApJ, 792, 53

\bibitem[\protect\citeauthoryear{Vasiliev}{2014}]{Vasiliev:2014}
Vasiliev, E., 2014, Class. Quantum Gravity, 31, 244002
\bibitem[\protect\citeauthoryear{Vasiliev \& Merritt}{2013}]{Vasiliev:2013}
Vasiliev, E., Merritt, D., 2013, ApJ, 774, 87

\bibitem[\protect\citeauthoryear{Vasudevan \& Fabian}{2007}]{Vasudevan:2007}
Vasudevan, R. V.,  Fabian, A. C., 2007, MNRAS, 381, 1235

\bibitem[\protect\citeauthoryear{Vinko et al.}{2015}]{Vinko:2015}
Vinko, J., Yuan, F., Quimby, R. M., Wheeler, J. C., Ramirez-Ruiz, E., et al., 2015, ApJ, 798, 12

\bibitem[\protect\citeauthoryear{Wang \& Merritt}{2004}]{Wang:2004}
Wang, J., Merritt, D., 2004, ApJ, 600, 149

\bibitem[\protect\citeauthoryear{Wang et al.}{2012}]{Wang:2012}
Wang, T.,  et al., 2012, ApJ, 749, 115


\bibitem[\protect\citeauthoryear{Wegg \& Bode}{2011}]{Wegg:2011}
Wegg, C., Bode, J. N., 2011, ApJ, 738, 8

\bibitem[\protect\citeauthoryear{Weisskopf}{2015}]{Weisskopf:2015}
Weisskopf, M. C., Gaskin, J.,  Tananbaum, H., Vikhlinin, arXiv:1505.00814

\bibitem[\protect\citeauthoryear{Wyithe \& Loeb}{2013}]{Wyithe:2013}
Wyithe, S., Loeb, A., 2013, MNRAS, 428, 2741

\bibitem[\protect\citeauthoryear{Woods \& Loeb}{1998}]{Woods:1998}
Woods, E., Loeb, A., 1998, ApJ 508, 760

\bibitem[\protect\citeauthoryear{Yuan et al.}{2014}]{Yuan:2014}
Yuan, W., Zhou, H., Dou, L., et al. 2014, ApJ, 782, 55

\bibitem[\protect\citeauthoryear{Zauderer et al.}{2011}]{Zauderer:2011}
Zauderer, B. A., Berger, E., Soderberg, A. M., et al., 2011, Nature, 476, 425


\bibitem[\protect\citeauthoryear{Zauderer et al.}{2013}]{Zauderer:2013}
Zauderer, B. A., Berger, E., Margutti, R., et al., 2013, Nature, 476, 425

\label{lastpage}

\end{thebibliography}
\end{document}